\documentclass[5p,times]{elsarticle}

\usepackage{amssymb}
\usepackage{graphicx}
\usepackage{subcaption}
\usepackage{array}
\usepackage{multirow}
\usepackage{algorithm}
\usepackage{algpseudocode}
\usepackage{amsmath}
\usepackage{booktabs}
\usepackage{tablefootnote}
\usepackage{listings}
\usepackage{xcolor}
\usepackage{enumitem}
\usepackage[switch]{lineno}
\usepackage{tabularx}
\usepackage[hidelinks]{hyperref}
\usepackage{pgfplots}
\pgfplotsset{compat=1.18}

\journal{Elsevier}

\begin{document}

\begin{frontmatter}



\title{DMLDroid: Deep Multimodal Fusion Framework for Android Malware Detection with Resilience to Code Obfuscation and Adversarial Perturbations
}

\author[inst1,inst2,inst3]{Doan Minh Trung}\ead{trungdm@uit.edu.vn}
\author[inst1,inst2]{Tien Duc Anh Hao}\ead{22520404@gm.uit.edu.vn}
\author[inst1,inst2]{Luong Hoang Minh}\ead{22520868@gm.uit.edu.vn}
\author[inst1,inst2,inst3]{\\Nghi Hoang Khoa}\ead{khoanh@uit.edu.vn}
\author[inst1,inst2]{Nguyen Tan Cam}\ead{camnt@uit.edu.vn}
\author[inst1,inst2,inst3]{Van-Hau Pham}\ead{haupv@uit.edu.vn}
\author[inst1,inst2,inst3]{Phan The Duy \corref{cor1}}\ead{duypt@uit.edu.vn}
\cortext[cor1]{Corresponding author.}

\affiliation[inst1]{organization={Information Security Laboratory, University of Information Technology},
            city={Ho Chi Minh city},
            country={Vietnam}}

\affiliation[inst2]{organization={Vietnam National University Ho Chi Minh City},
            city={Ho Chi Minh City},
            country={Vietnam}}

\affiliation[3]{organization={VNU-HCM Information Security Center},
    city={Ho Chi Minh City},
    country={Vietnam}}
    
\begin{abstract}

In recent years, learning-based Android malware detection has seen significant advancements, with detectors generally falling into three categories: string-based, image-based, and graph-based approaches. While these methods have shown strong detection performance, they often struggle to sustain robustness in real-world settings, particularly when facing code obfuscation and adversarial examples (AEs). Deep multimodal learning has emerged as a promising solution, leveraging the strengths of multiple feature types to enhance robustness and generalization. However, a systematic investigation of multimodal fusion for both accuracy and resilience remains underexplored. In this study, we propose DMLDroid, an Android malware detection based on multimodal fusion that leverages three different representations of malware features, including permissions \& intents (tabular-based), DEX file representations (image-based), and API calls (graph-derived sequence-based). We conduct exhaustive experiments independently on each feature, as well as in combination, using different fusion strategies. Experimental results on the CICMalDroid 2020 dataset demonstrate that our multimodal approach with the dynamic weighted fusion mechanism achieves high performance, reaching 97.98\% accuracy and 98.67\% F1-score on original malware detection. Notably, the proposed method maintains strong robustness, sustaining over 98\% accuracy and 98\% F1-score under both obfuscation and adversarial attack scenarios. Our findings highlight the benefits of multimodal fusion in improving both detection accuracy and robustness against evolving Android malware threats.

\end{abstract}




\end{frontmatter}


\section{Introduction} \label{sec_introduction}
Android malware has become a major cybersecurity threat as the Android platform continues to grow, with over three billion active users across 190 countries \cite{curry2025android}. In recent years, numerous learning-based methods have been developed to detect and classify Android malware using dynamic, static, and hybrid analysis \cite{liu2022deep, gopinath2023comprehensive}. However, dynamic analysis is typically time-consuming, resource-intensive, and dependent on emulator environments. Consequently, most detection methods rely on static features, which enable faster and more efficient detection without executing the applications.

Static analysis techniques can be grouped into three categories: string-based, image-based, and graph-based. String-based methods extract sequences of important strings, such as permissions and API calls, from APK files. These sequences are then transformed into feature vectors for malware detection. For instance, work proposed in \cite{elayan2021android} leveraged API calls and permissions with a GRU-based deep learning model, achieving 98.2\% accuracy on the CICAndMal2017 dataset. Image-based methods convert APK files into visual representations and apply image classification techniques to distinguish between benign and malicious applications. V. Ravi et al. \cite{ravi2023efficientnet} explored 26 CNN-based pre-trained models for Android malware detection, extracting image features and utilizing a fusion-based classification approach. Their EfficientNet-based fused models outperformed other CNN variants, demonstrating strong generalization across multiple datasets. Graph-based methods capture structural and semantic relationships within an APK by constructing various graphs, such as function call graph (FCG), control flow graph (CFG), and API call graph (ACG). These graphs are then transformed into feature vectors and processed by machine learning (ML) and deep learning (DL) models for classification. Recent studies highlight the effectiveness of Graph Neural Networks (GNNs) in learning expressive embeddings from malware graphs like FCGs and CFGs, achieving state-of-the-art detection performance \cite{bilot2024survey}. Compared to traditional ML and DL techniques, GNN-based methods better capture complex dependencies in Android applications, making them a promising approach for malware detection. Additionally, while the integration of Large Language Models (LLMs) with GNNs has shown promise in various graph learning tasks \cite{jin2024large}, its potential for Android malware detection remains an open research direction. This synergy could enhance feature extraction and improve GNNs' ability to capture intricate dependencies, paving the way for more robust detection methods in future research.

While string-based, image-based, and graph-based methods have each shown strong detection capabilities, they still struggle to maintain performance when facing out-of-distribution (OOD) samples. Such OOD samples are commonly introduced through adversarial examples (AEs) and code obfuscation techniques. Gao et al. \cite{gao2024comprehensive} conducted a comprehensive study of 12 representative detectors across these three categories and found that none of them could sustain their ideal-setting detection performance under such challenging conditions. 

To address these vulnerabilities, prior research has explored robustness-enhancing strategies such as adversarial training, regularization methods, feature manipulation, and ensemble approaches, with a recent survey by Bhusal et al. \cite{bhusal2025adversarial} providing an in-depth overview of their effectiveness across different domains. In parallel, efforts to counter code obfuscation have also emerged, for instance by leveraging Markov image representations with CNN-based classifiers \cite{dhanya2023obfuscated} or multimodal with a voting fusion mechanism that combines Function Call Graphs and Opcode-based Markov transition matrices \cite{gao2023obfuscation}. Building on these lines of work, the resurgence of multimodal learning has shown great promise for Android malware detection \cite{de2023chimera, li2025detecting, zhang2025mpdroid, li2025multimodal}, as combining complementary modalities often leads to stronger detection performance and improved generalization compared to single-modality baselines. 

Despite these advances, existing multimodal approaches have fundamental limitations. To date, no study has systematically evaluated resilience against both adversarial attacks and obfuscation, leaving their effectiveness in real-world scenarios uncertain. For instance, X. Li et al. \cite{li2025detecting} integrate programming language and machine language modalities as texts and RGB images, using a multi-head self-attention fusion mechanism. However, converting raw bytes into RGB images is fragile under code obfuscation \cite{gao2024comprehensive}, and the study lacks systematic evaluation under adversarial and obfuscation scenarios, leaving it unclear whether the fusion mechanism can mitigate corrupted modalities. Similarly, Oliveira et al. \cite{de2023chimera} fuse features such as permissions, intents, system calls, and grayscale DEX images through simple concatenation. Permissions and intents are especially vulnerable to adversarial perturbations like GAN-based attacks \cite{rathore2021robust}, while grayscale DEX images are prone to obfuscation. The simple concatenation strategy can also introduce a unimodal bias, causing the model to rely mainly on a single modality. Gao et al. \cite{gao2023obfuscation} explicitly target obfuscation resilience with a voting-based fusion mechanism, where predictions from multiple classifiers are combined using individual confidence-based weights. This design can help balance the weaknesses of individual classifiers and improve robustness against obfuscation. However, the study does not consider adversarial robustness, leaving open the question of whether their voting-based fusion can withstand maliciously crafted inputs. Collectively, these limitations highlight the absence of a purposefully designed multimodal fusion framework capable of simultaneously mitigating both obfuscation and adversarial perturbations.

Motivated by this gap, we propose DMLDroid, a resilient Android malware detection system based on deep multimodal fusion. Our framework integrates features from different modalities at the intermediate level and jointly feeds them into the model for decision making. We leverage three common types of malware features: a multilayer perceptron (MLP) models permissions and intents as tabular data, a convolutional neural network (CNN) analyzes RGB images derived from DEX files, and DistilBERT encodes API call graph sequences to capture contextual dependencies. At the fusion stage, we investigate multiple strategies, including simple concatenation, self-attention, cross-attention, gated fusion, and dynamic weighted fusion, which allows us to systematically assess their impact on performance and robustness. By combining heterogeneous feature representations within these fusion frameworks, DMLDroid captures complementary information across modalities, improves generalization, and strengthens resilience against both adversarial attacks and obfuscation techniques.

The main contributions of this work are summarized as follows:
\begin{itemize}
    \item We propose a resilient Android malware detection framework based on deep multimodal fusion, explicitly designed to address both adversarial perturbations and code obfuscation by combining tabular (permissions \& intents), image-based (DEX RGB), and sequential (API graph sequences) representations.
    \item We provide the first systematic comparison of five intermediate fusion strategies—concatenation, self-attention, cross-attention, gated fusion, and dynamic weighted fusion—highlighting their relative strengths and limitations under adversarial and obfuscation scenarios.
    \item We design extensive experimental settings that incorporate real-world evasion challenges, including GAN-based adversarial example generation and multiple obfuscation categories, offering a more realistic robustness evaluation than prior multimodal works.
    \item We benchmark our approach against state-of-the-art multimodal baselines on the CICMalDroid 2020 dataset, achieving superior performance in terms of both detection accuracy and robustness, while also analyzing the trade-off between detection effectiveness and computational cost.
\end{itemize}

The remainder of this paper is organized as follows: \textbf{Section \ref{sec_relatedworks}} reviews related work in Android malware detection, adversarial and obfuscation robustness. \textbf{Section \ref{sec_design}} describes our proposed approach, detailing the feature extraction process, model architecture, and fusion strategies. \textbf{Section \ref{sec_experiment}} presents the experimental setup, including datasets, evaluation metrics, adversarial attack, and obfuscation scenarios. \textbf{Section \ref{sec_threat_to_validity}} discusses the threat factors that impact the validity of results and insights gained from our experiments. Finally, \textbf{Section \ref{sec_conclusion}} concludes the paper with a summary of findings and future research directions.

\section{Related Works} \label{sec_relatedworks}
Previous studies on Android malware detection have addressed robustness from different perspectives, particularly in the face of code obfuscation and adversarial attacks. \textbf{Table \ref{tab:sota-reviews}} provides a concise overview of these key works, highlighting their methodologies and contributions to the field.

\begin{table*}[!ht]
\centering
\small
\caption{Highlights of reviewed related works in Android malware detection.}
\label{tab:sota-reviews}
\begin{tabular}{|c|c|c|c|c|c|c|}
\hline
\textbf{Ref.} &
  \textbf{Dataset} &
  \textbf{Input/features} &
  \textbf{Classifiers} &
  \textbf{Performance} &
  \textbf{\begin{tabular}[c]{@{}c@{}}Obfuscation\\ resilience\end{tabular}} &
  \textbf{\begin{tabular}[c]{@{}c@{}}Adversarial\\ resilience\end{tabular}} \\ \hline
Bai's (2021) \cite{bai2021comparative} &
  \begin{tabular}[c]{@{}c@{}}Genome, Drebin, \\ AMD\end{tabular} &
  \begin{tabular}[c]{@{}c@{}}Permissions,\\  API calls, ICC\end{tabular} &
  MLP &
  Acc = 99.20\% &
  Yes &
  No \\ \hline
\begin{tabular}[c]{@{}c@{}}R. Yumlembam\\ (2022) \cite{yumlembam2022iot} \end{tabular} &
  \begin{tabular}[c]{@{}c@{}}CICMalDroid\\ 2020\end{tabular} &
  API call graph &
  GraphSage, CNN &
  Acc = 98.33\% &
  N/A &
  No \\ \hline
\begin{tabular}[c]{@{}c@{}}KA. Dhanya\\ (2023) \cite{dhanya2023obfuscated} \end{tabular} &
  \begin{tabular}[c]{@{}c@{}}Drebin, \\ CICInvesAnd-\\ Mal2019, \\ GooglePlay\end{tabular} &
  \begin{tabular}[c]{@{}c@{}}Markov image \\ from DEX file\end{tabular} &
  CNN &
  Acc = 99.41\% &
  Yes &
  N/A \\ \hline
\begin{tabular}[c]{@{}c@{}}Y. He (2022)\\ \cite{he2022msdroid} \end{tabular} &
  \begin{tabular}[c]{@{}c@{}}Drebin, AndroZoo, \\ AMD\end{tabular} &
  \begin{tabular}[c]{@{}c@{}}API call graph, \\ permissions, \\ opcodes\end{tabular} &
  GNN &
  Acc = 97.82\% &
  Yes &
  N/A \\ \hline
\begin{tabular}[c]{@{}c@{}}C. Gao (2023)\\ \cite{gao2023obfuscation} \end{tabular} &
  AndroZoo, Drebin &
  \begin{tabular}[c]{@{}c@{}}API call graph, \\ Dalvik instructions \\ sequence\end{tabular} &
  \begin{tabular}[c]{@{}c@{}}Multimodal fusion\\ model (GCN, \\ EfficientNet)\end{tabular} &
  F1 = 97.90\% &
  Yes &
  N/A \\ \hline
\begin{tabular}[c]{@{}c@{}}AS de Oliveira\\ (2023) \cite{de2023chimera} \end{tabular} &
  Omnidroid &
  \begin{tabular}[c]{@{}c@{}}Intents, permissions, \\ DEX grayscale \\ image, system call \\ sequences\end{tabular} &
  \begin{tabular}[c]{@{}c@{}}Multimodal fusion\\ model (DNN, \\ CNN, Transformer \\ Networks)\end{tabular} &
  Acc = 90.90\% &
  N/A &
  N/A \\ \hline
\begin{tabular}[c]{@{}c@{}}X. Li (2025)\\ \cite{li2025detecting} \end{tabular} &
  \begin{tabular}[c]{@{}c@{}}VirusShare, \\ GooglePlay\end{tabular} &
  \begin{tabular}[c]{@{}c@{}}DEX RGB image,\\ API call graph\end{tabular} &
  \begin{tabular}[c]{@{}c@{}}Multimodal fusion\\ model (ViT,\\ UniXcoder)\end{tabular} &
  Acc = 98.28\% &
  N/A &
  N/A \\ \hline
\begin{tabular}[c]{@{}c@{}}\textbf{DMLDroid} \\ \textbf{(Ours)}\end{tabular} &
  \textbf{\begin{tabular}[c]{@{}c@{}}CICMalDroid \\ 2020\end{tabular}} &
  \textbf{\begin{tabular}[c]{@{}c@{}}Permission \\ \& Intents, \\ DEX RGB image, \\ API call sequences\end{tabular}} &
  \textbf{\begin{tabular}[c]{@{}c@{}}Multimodal fusion\\ model (MLP, \\ CNN, DistilBERT)\end{tabular}} &
  \textbf{Acc = 97.98\%} &
  \textbf{Yes} &
  \textbf{Yes} \\ \hline
\end{tabular}
\end{table*}

\subsection{Robustness against obfuscation}

Obfuscation techniques, such as code encryption, method renaming, and the insertion of junk code, have been shown to degrade the performance of static-analysis-based detectors. In fact, over 95\% of obfuscation techniques directly target the \textit{classes.dex} file, making detection models that rely solely on this source less reliable \cite{gao2024comprehensive}. Gao et al. \cite{gao2024comprehensive} further reproduced image-based approaches that convert \textit{classes.dex} into color images \cite{xiao2019image} or Markov images \cite{yuan2020byte} and reported that these methods were significantly affected by obfuscation. Interestingly, Dhanya \cite{dhanya2023obfuscated} adopted a similar strategy by converting \textit{classes.dex} into Markov images and training a CNN. Their approach, however, achieved a remarkably high accuracy of 99.65\% in detecting obfuscated Android malware in IoT environments. This discrepancy highlights that while \textit{classes.dex} features are generally vulnerable to obfuscation, the effectiveness of detection can also depend on how these features are represented and modeled. 

Beyond representation learning, another effective direction for mitigating obfuscation is the integration of diverse feature sources. Extracting complementary information from components such as manifest files, native libraries, or dynamic behaviors has been suggested to improve robustness. For instance, Bai et al. \cite{bai2021comparative} constructed a dataset comprising 250 widely used manual syntactic features, including permissions, API calls, and attributes of inter-component communication (ICC), and trained several classical ML models. Their results demonstrated that combining multiple feature categories consistently outperformed single-feature approaches, and such combinations have also been shown to offer stronger resilience against obfuscation \cite{gao2024comprehensive}. Similarly, Y. He et al. \cite{he2022msdroid} proposed MSDROID, a snippet-based detector leveraging GNNs with a combination of API call graph, opcode, and permission features. Extensive experiments confirmed its effectiveness and robustness against obfuscation, surpassing existing methods. 

Recent works have explored multimodal learning in malware detection, integrating heterogeneous features via various fusion strategies to enhance resilience against obfuscation. For example, Gao et al. \cite{gao2023obfuscation} proposed CorDroid, which combines sensitive function call graphs with opcode-based Markov transition matrices to capture Android app behaviors from multiple perspectives. Their method achieves higher accuracy, robustness, and efficiency compared to state-of-the-art (SOTA) approaches, even on obfuscated datasets. In \cite{de2023chimera}, Oliveira et al. introduced Chimera, a multimodal deep learning method for Android malware detection. Their approach fuses heterogeneous features, including DEX file images, intents \& permissions, and system call sequences, which are then processed by a deep neural architecture to perform classification. Similarly, Li et al. \cite{li2025detecting} leveraged large pre-trained models with a fine-grained multimodal fusion strategy that integrates source code and binary code as programming language and machine language modalities, respectively. Both studies report strong performance on unseen data. However, their robustness against complex obfuscation techniques has not been rigorously evaluated.

\subsection{Robustness against adversarial attacks}
Adversarial examples have emerged as a critical threat to ML/DL-based malware detection, as they can subtly manipulate inputs to bypass detection systems and severely degrade model accuracy. Among these, GAN-based adversarial attacks are particularly concerning due to their ability to generate diverse and realistic variants that effectively exploit model vulnerabilities. Originating from the pioneer work MalGAN \cite{hu2022generating}, which demonstrated black-box evasion against Android malware detectors, subsequent research has developed numerous GAN variants tailored to Android malware. For instance, E-MalGAN \cite{li2019adversarial} focuses on generating evasive samples to bypass cloud-based detectors. Others manipulate specific low-level features, like AndrOpGAN \cite{zhang2020andropgan}, which modifies opcode distributions, or VGAE-MalGAN \cite{yumlembam2022iot}, which alters API graphs to deceive GNN-based classifiers while preserving semantics. Furthermore, improvements to GAN stability have been explored, with LSGAN \cite{wang2021lsgan} applying a least square loss function and a modified CycleGAN \cite{atedjio2024cyclegan} using a gradient penalty to enhance model training and detection rates.

Most current adversarial defense and detection works primarily rely on adversarial training and retraining to enhance model robustness \cite{gebrehans2025generative}. In \cite{yumlembam2022iot}, the authors used the generative model VGAE-MalGAN to create AEs. These attacks caused their proposed GraphSage + CNN malware detection model to suffer a significant accuracy drop from 98.6\% to just 47.72\% on the CICMalDroid 2020 dataset. To address this, they performed retraining using these AEs, which successfully improved the model's accuracy to 97.86\%. Similarly, in \cite{wang2021lsgan}, traditional detection models such as Decision Tree (DT), Logistic Regression (LR), Multilayer Perceptron (MLP), and K-Nearest Neighbors (KNN) all demonstrated a significant drop in performance when attacked by AEs generated by the LSGAN model. To address this, the authors also employed adversarial retraining as their primary defense. However, this retraining process is often computationally intensive and time-consuming, making it unfeasible for real-time systems that demand rapid and continuous updates.

Multimodal learning has shown great potential in mitigating obfuscation and can also enhance robustness against adversarial attacks. However, a notable research gap exists, as no current studies have rigorously evaluated multimodal Android malware detectors against adversarial attacks. Existing works also have limitations in assessing complex obfuscation techniques. Our study addresses this gap by comprehensively evaluating the robustness of our proposed multimodal approach against both types of attacks.

\section{Design of proposed approach} \label{sec_design}
\subsection{Overview of DMLDroid}
\begin{figure*}[!ht]
    \centering
    \includegraphics[width=0.95\textwidth]{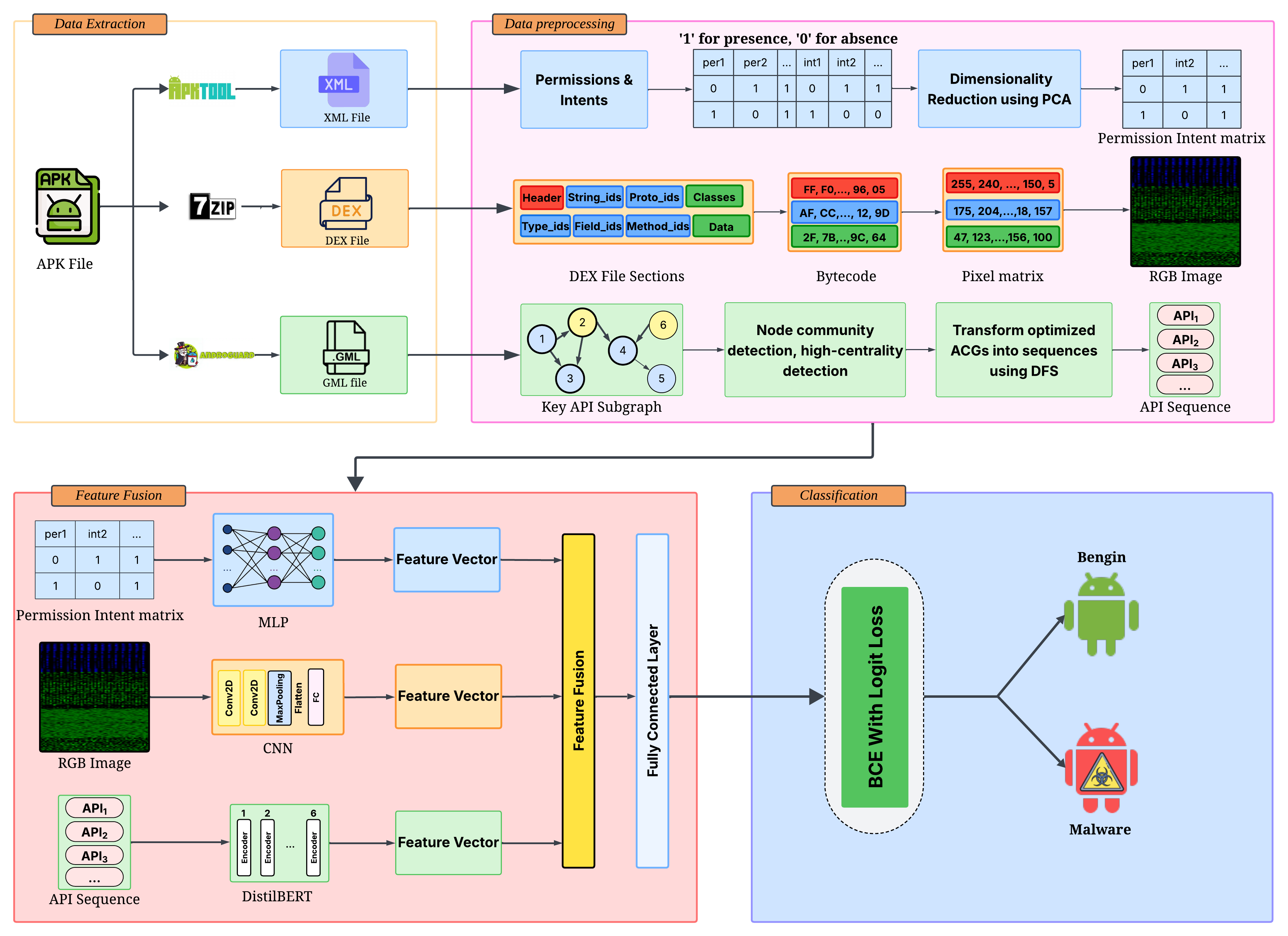}
    \caption{The overall architecture of DMLDroid.}
    \label{fig:arch_dmlDroid}
\end{figure*}

\textbf{Figure \ref{fig:arch_dmlDroid}} illustrates our proposed framework, namely DMLDroid, which consists of two main modules: feature representation and the feature fusion \& classification module. We chose permissions \& intents, DEX file section features, and API calls since they have been the dominant feature types in nearly all prior Android malware studies. MLPs and CNNs were employed as they are widely adopted in this domain \cite{liu2022deep}. At the same time, pre-trained language models were further explored for API sequence analysis \cite{saracino2023graph}, given their proven potential in effectively detecting malware through contextual and long-range dependency modeling.

The first step involves extracting, preprocessing, and representing static features across three modalities: tabular-based, image-based, and graph-derived sequence-based features. Tabular-based features (TF) are obtained by decompiling the APK with Apktool \cite{iBotPeaches} to extract permissions and intents from AndroidManifest.xml, capturing security-sensitive declarations. We apply PCA to reduce dimensionality while preserving essential information, and an MLP further processes these features to generate high-level representations. In the image-based features (IF), the APK file is unzipped to extract the DEX file, which contains the compiled bytecode of the application. The data section of the DEX file, including variable names, method references, and class definitions, is then structured and converted into an RGB image, where byte values are mapped to pixel intensities. A CNN is then utilized to extract feature representations from these images. For graph-derived sequence-based features (GSF), Androguard, a well-known tool,  is utilized to generate ACG from Android APKs. Given the substantial variation in ACG sizes, with node counts ranging from thousands to millions, we construct an optimized graph reduction method. This method leverages community detection and centrality measures to identify key components while removing irrelevant API nodes. The objective is to minimize and standardize the graph’s scale while preserving its essential features and structural integrity. Subsequently, a pre-trained DistilBERT model extracts feature representations from the refined graph data.

In the feature fusion \& classification module, we experiment with both intermediate and late fusion strategies to assess their impact on classification performance. Intermediate Fusion integrates features from all three modalities to construct a unified multimodal representation of application behavior, which is then fed into a linear classifier for final predictions. In contrast, late fusion combines the classification outputs from individual modalities using an ensemble-based approach. Through these experiments, we evaluate the effectiveness of each method in capturing cross-modal correlations and improving detection accuracy.

\subsection{Feature representation of Android application modality}
\subsubsection{Tabular-based feature representation} \label{sec:tab-rep}

\begin{figure*}[!ht]
    \centering
    \includegraphics[width=1\linewidth]{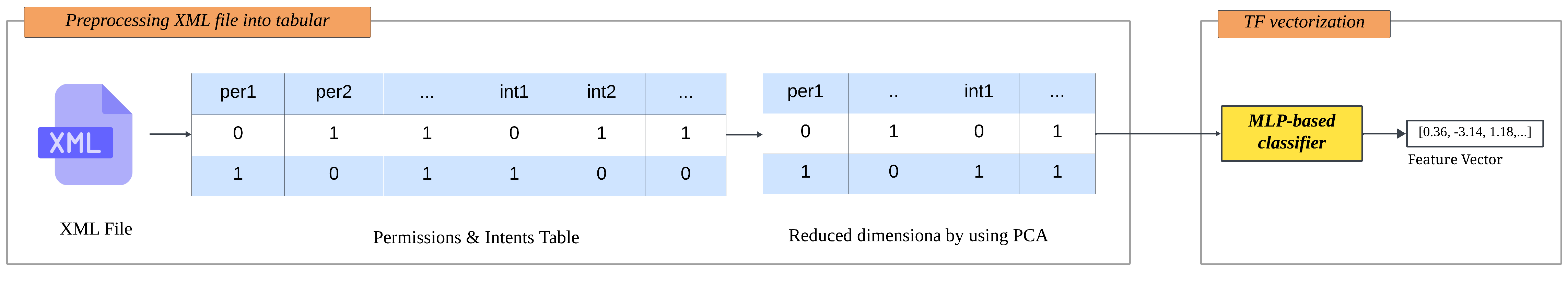}
    \caption{Tabular-based feature representation.}
    \label{fig:tab-repre}
\end{figure*}

According to the Android SDK (Software Development Kit), an application defines its functionalities in the manifest file. By analyzing permissions and intents, it is possible to extract high-level semantic information that helps distinguish benign applications from malware. These features serve as crucial inputs for ML and DL models, enhancing their ability to learn discriminative patterns and improve malware detection accuracy \cite{imtiaz2021deepamd, sharma2024ipanalyzer}. \textbf{Figure \ref{fig:tab-repre}} provides an overview of the entire process.

\textbf{a) Preprocessing.} After parsing the manifest file, the extracted features include permissions, actions, services, and categories, as illustrated in \textbf{Table \ref{tab:pi-fea-example}}. These features are structured in a CSV format. Each APK \( A_i \) is represented as \( A_i = (H_i, x_{i1}, x_{i2}, \dots, x_{in}) \) where \( H_i \) denotes the SHA-256 hash for unique identification, and each binary feature \( x_{ij} \in \{0,1\} \) indicates the presence ('1') or absence ('0') of feature \( f_j \). 
The dataset forms a binary matrix \( \mathbf{X}_{tab} \in \{0,1\}^{m \times n} \), where rows correspond to APKs and columns to features.

\begin{table}[!ht]
\centering
\small
\caption{Example of permissions and intent features.}
\begin{tabular}{|l|l|}
\hline
\textbf{Type}                     & \textbf{Feature}         \\ \hline
\multirow{4}{*}{Permissions}      & ACCESS\_COARSE\_LOCATION \\ \cline{2-2} 
                                  & READ\_EXTERNAL\_STORAGE  \\ \cline{2-2} 
                                  & INTERNET                 \\ \cline{2-2} 
                                  & CAMERA                   \\ \hline
\multirow{4}{*}{Intents action}   & ACTION\_BOOT\_COMPLETED  \\ \cline{2-2} 
                                  & PHONE\_STATE             \\ \cline{2-2} 
                                  & SMS\_RECEIVED            \\ \cline{2-2} 
                                  & MEDIA\_MOUNTED           \\ \hline
\multirow{4}{*}{Services}         & FirebaseMessagingService \\ \cline{2-2} 
                                  & KeepAliveService         \\ \cline{2-2} 
                                  & LocationRetrievalService \\ \cline{2-2} 
                                  & WalletApduService        \\ \hline
\multirow{4}{*}{Intents category} & default                  \\ \cline{2-2} 
                                  & MULTIWINDOW\_LAUNCHER    \\ \cline{2-2} 
                                  & LAUNCHER\_APP            \\ \cline{2-2} 
                                  & BROWSABLE                \\ \hline
\end{tabular}
\label{tab:pi-fea-example}
\end{table}

As the dataset grows, the increasing feature space leads to high-dimensional data, which degrades model performance and increases computational overhead. To address this, principal component analysis (PCA) is applied to \( \mathbf{X}_{tab} \), reducing redundancy while preserving 90\% of the variance \cite{csahin2021permission}. The transformed dataset \( \mathbf{P}_{tab} \in \mathbb{R}^{m \times d} \), where \( d \ll n \), retains essential latent features: 
\(
\mathbf{P}_{tab} = \{ p_1, p_2, \dots, p_m \}, \quad p_i = (p_{i1}, p_{i2}, \dots, p_{id})
\). Each \( p_i \) is a compact representation of APK \( A_i \), forming the input for the next transformation stage.

\textbf{b) TF vectorization.} Following dimensionality reduction, an MLP is employed to vectorize the transformed dataset, excluding \( H_i \), to obtain the TF feature dataset. 
The MLP progressively extracts abstract representations of the TF features, capturing essential patterns while reducing redundancy. The final hidden layer outputs a compact TF feature vector that effectively characterizes the APK. Formally, the TF feature dataset is defined as: \(
\mathbf{TF\_dataset} = \{ v_1^{TF}, v_2^{TF}, \dots, v_m^{TF} \} \), where \( v_i^{TF} \in \mathbb{R}^{1 \times 128} \)  denotes the TF feature vector of the \( i \)-th APK.

\subsubsection{Image-based feature representation}
\begin{figure*}[!ht]
    \centering
    \includegraphics[width=1\linewidth]{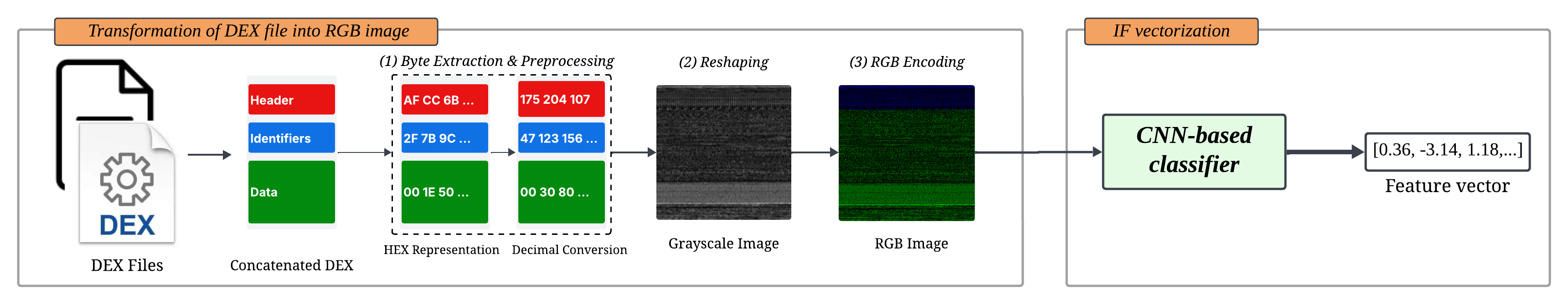}
    \caption{Image-based feature representation.}
    \label{fig:dex_rgb}
\end{figure*}

While tabular-based analysis relies on binary '0' and '1' representations, making it susceptible to adversarial attacks \cite{li2019adversarial, rathore2021robust}, recent studies have explored alternative approaches to enhance robustness. One promising direction is transforming DEX files into RGB image representations, which has been shown to improve the accuracy of Android malware detection \cite{darwaish2020rgb, fang2020android}. However, existing image conversion techniques, which typically map consecutive bytecode sequences to RGB pixels, exhibit significant brittleness against obfuscation \cite{gao2024comprehensive}. Their reliance on a strict byte order makes them highly sensitive to syntactic changes like dead-code insertion, which can destroy essential visual patterns. To address this limitation, our approach instead encodes the semantic structure of the DEX file by mapping its core sections (header, class definitions/identifiers, and data) to the R, G, and B channels, respectively. This method preserves the file's macro-level layout, creating a stable visual signature that remains consistent even when the content within sections is altered by obfuscation. \textbf{Figure \ref{fig:dex_rgb}} illustrates this process in details.

\textbf{a) Transformation of DEX file into RGB image.} For a given APK file \( A_i \), we extract its contents to obtain a set of DEX files, denoted as \( DEX_{A_i} = \{ \textit{dex}_1, \textit{dex}_2, ..., \textit{dex}_k \} \), where each \(\textit{dex}_j\) represents an individual DEX file. If an APK contains multiple DEX files due to the 65K method limit, we adopt the multidex approach proposed by Fang et al. \cite{fang2020android}. Since all DEX files share a similar structure, corresponding sections such as headers, identifiers, and data are extracted separately and then concatenated to maintain structural integrity while enabling a unified representation for visualization. To transform these DEX files into RGB images, we follow a structured approach:

\textit{(1) Byte extraction and preprocessing:} Each DEX file \(\textit{dex}_j\) is converted into its hexadecimal representation, forming a one-dimensional (1D) byte array \( \mathbf{B}_{j} = (b_1, b_2, ..., b_l) \) where \( b_i \in [0,255] \). 

\textit{(2) Reshaping into a grayscale image:} The 1D byte array \( \mathbf{B}_{j} \) is reshaped into a two-dimensional (2D) grayscale image as \( \mathbf{G}_j = \text{reshape}(\mathbf{B}_j, h, w) \), where \( w \) is a fixed width and height \( h = l/w \) is the computed height based on the total byte length \( l \).

\textit{(3) RGB channel encoding:} The grayscale image \( \mathbf{G}_j \) is mapped to an RGB image \( \mathbf{I}_j \) by assigning byte values to different color channels based on their positions in the DEX file. The structure of a DEX file consists of headers, identifiers, class definitions, and data. The bytes in the header section are assigned to the red color channel,  those from the identifier and class definition regions to the green channel, and data section bytes to the blue channel. As a result, a layer of semantic encoding is added to the image, which helps analyze the Android malware more effectively.

After processing all the APK files in the dataset, we construct a corresponding collection of DEX files, denoted as
\(
\mathcal{D}_{dex} = \{DEX_{A_1}, DEX_{A_2}, ..., DEX_{A_N} \}
\) where each \( DEX_{A_i} \) represents the set of DEX files extracted from APK \( A_i \). We then generate an image dataset corresponding to these DEX files, denoted as
\(
\mathcal{D}_{img} = \{ I_1, I_2, ..., I_M \}
\)
where each \( I_j \) is an RGB image transformed from a DEX file \( dex_j \in \mathcal{D}_{dex} \). 

\textbf{b) IF vectorization.} Following the transformation of DEX files into RGB images, a CNN is employed to extract high-level image-based features, forming the IF feature dataset. The CNN captures spatial patterns and hierarchical representations within the images, highlighting structural and semantic relationships of the DEX files. The final hidden layer produces a compact IF feature vector that effectively represents the APK’s image-based characteristics. Formally, the IF feature dataset is defined as: \(
\mathbf{IF\_dataset} = \{ v_1^{IF}, v_2^{IF}, \dots, v_m^{IF} \} \), where \( v_i^{IF} \in \mathbb{R}^{1 \times 128} \) denotes the IF feature vector of the \( j \)-th image \( I_j \).

\subsubsection{Graph-based feature representation}
\begin{figure*}[!ht]
    \centering
    \includegraphics[width=1\linewidth]{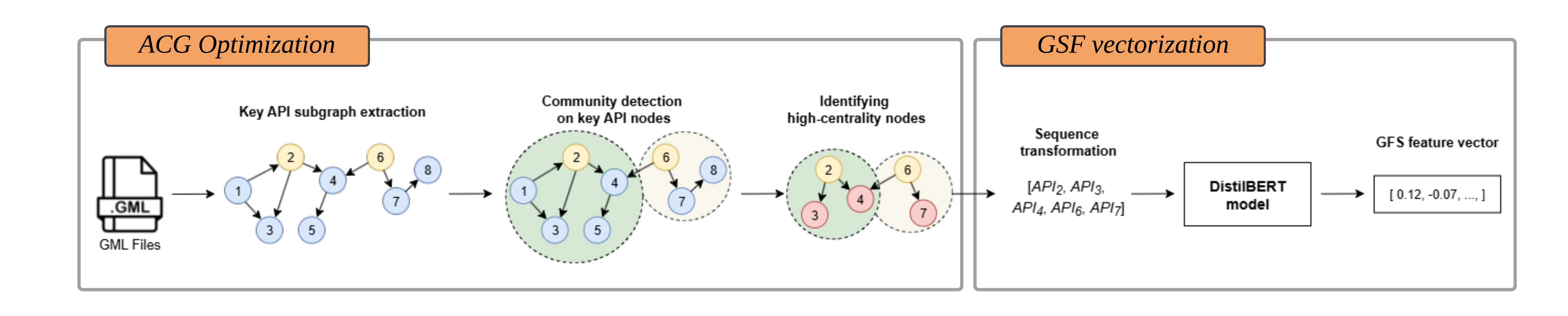}
    \caption{Graph-based feature representation.}
    \label{fig:gsf-repr}
\end{figure*}

Because Android apps use a large number of APIs, graph-based methods often face scalability challenges, requiring substantial computational resources to handle the complexity of API call graphs. Previous studies \cite{gao2021gdroid, shen2024ghgdroid, li2025detecting} have addressed this by leveraging sensitive API sets from tools like PScout \cite{au2012pscout} and Susi \cite{arzt2013susi} to reduce graph complexity while maintaining key structural information for malware detection. However, these API sets have become outdated, as many APIs in the datasets used in our study are no longer relevant. As a result, these methods may lose effectiveness over time, especially given the continuous evolution of Android APIs and the increasingly sophisticated nature of malware. Therefore, we propose an optimized graph reduction method that dynamically extracts key APIs from the dataset used, constructing a feature representation that enhances adaptability to evolving malware behaviors while reducing reliance on predefined API sets. The details of this process are shown in \textbf{Figure \ref{fig:gsf-repr}}.

\begin{figure}[!ht]
    \centering
    \includegraphics[width=1\linewidth]{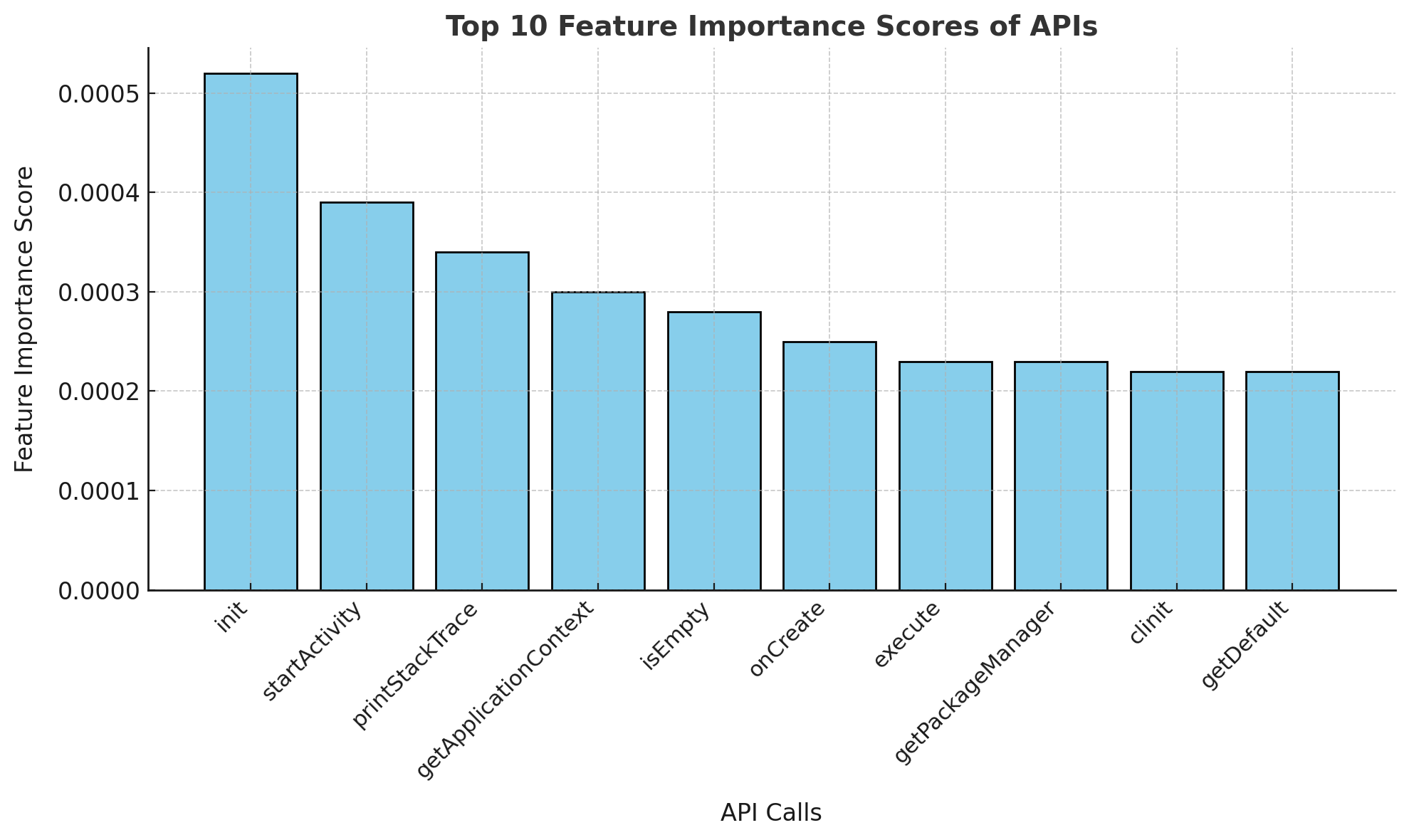}
    \caption{Most common key APIs with high feature importance.}
    \label{fig:common-key-apis}
\end{figure}

\textbf{a) ACG optimization.} Before implementing ACG optimization, we conduct a feature selection process inspired by TF-IDF, a technique widely used in NLP. This method helps identify APIs most frequently used by benign and malicious applications.  After selecting these commonly used APIs, we refine the selection by leveraging a Random Forest (RF) model to evaluate feature importance scores. \textbf{Figure \ref{fig:common-key-apis}} presents the top-ranked APIs by importance scores. We select the APIs with the highest scores, which serve as pivotal transition points where benign and malicious applications diverge in execution. Based on this, the key API call set is defined as: \( key\ API\ set = \{ key\_api_1, key\_api_2, \dots, key\_api_m \} \) where each \( key\_api_{i} \) is a selected API call that meets the importance criteria based on the feature selection process. The workflow of the ACG optimization method is summarized as follows:

\textit{(1) Key API subgraph extraction:}  
Given a directed API call graph \( G = (V, E) \), we construct the key API subgraph \( G_k = (V_k, E_k) \) as follows:  \(
V_k = \{ v \in V \mid v \in \mathit{key\ API\ set} \}, \quad
E_k = \{ (u, v) \in E \mid u, v \in V_k \}
\) where \( V_k \) contains only nodes corresponding to the selected key APIs, and \( E_k \) includes edges representing direct interactions between them.

\textit{(2) Community detection on key API nodes:}  
While the key API subgraph \( G_k \) captures transition points between benign and malicious behaviors, it lacks structural context. Since APIs interact within functional groups, we apply community detection \cite{traag2019louvain} to identify clusters of key API nodes with strong intra-community connections, revealing underlying behavioral patterns.

Given \( G_k \), we define a community partition \( \mathcal{C} = \{ C_1, C_2, \dots, \\C_n \} \) such that each community \( C_i \) consists of strongly connected key API nodes. The communities are determined by optimizing modularity \( \mathcal{Q} \), defined as:

\begin{equation}
Q = \frac{1}{2|E_k|} \sum_{u,v \in V_k} \left[ A_{uv} - \gamma \frac{d_u d_v}{2|E_k|} \right] \delta(C_u, C_v)
\end{equation}

where \( A_{uv} \) is the adjacency matrix of \( G_k \); \( d_u \) and \( d_v \) are the degrees of nodes \( u \) and \( v \); \( \gamma \) is a resolution parameter controlling community granularity; \( \delta(C_u, C_v) = 1 \) if \( u \) and \( v \) belong to the same community, otherwise 0; and \( |E_k| \) is the total number of edges in \( G_k \). 

\textit{(3) Identifying high-centrality nodes:}
Using centrality measures, we identify the top neighbor nodes within the communities of key API nodes. These high-centrality nodes play a crucial role in API execution, influencing the information flow within and across communities.

Let \( V_k \) be the set of key API nodes. We define the neighbor node set \(V_n\) as: \(V_n = \{ v \in V \setminus V_k \mid (u, v) \in E, \ u \in V_k \} \), where \( V_n \) consists of all nodes that have a directed edge from at least one key API node in \( V_k \). We further define the subset of neighbor nodes within the same community as: \( V_n^c = \{ v \in V_n \mid v \in C_i, \ u \in C_i, \ \forall u \in V_k \} \), where \( V_n^c \) includes only those neighbor nodes that belong to the same community \( C_i \) as their corresponding key API nodes. To determine the most influential nodes among \( V_n \), we compute the following five centrality measures within each community:

\begin{itemize}
    \item Out-Degree Centrality (\( C_{out}(v) \)): Measures the number of outgoing edges from a node, indicating its influence over other nodes:
    \begin{equation}
    C_{out}(v) = \frac{\sum_{u \in C_i} A_{vu}}{|V_n^{C_i}| - 1}
    \end{equation}
    where \( A_{vu} \) is the adjacency matrix representing directed edges.
    \item Pagerank Centrality (\( C_{PR}(v) \)): Assigns importance to a node based on the number and quality of incoming edges within the community.
    \begin{equation}
    C_{\text{PR}}(v) = \alpha \sum_{u \in \mathcal{N}_{\text{in}}(v)} \frac{C_{\text{PR}}(u)}{d_{\text{out}}(u)} + (1 - \alpha) \frac{1}{|V_n^{C_i}|}
    \end{equation}
    where \( \mathcal{N}_{\text{in}}(v) \) is the set of nodes within the same community linking to \( v \), \( d_{\text{out}}(u) \) is the out-degree of the node \( u \), and \( \alpha \in (0,1) \) is the damping factor, typically set to 0.85, representing the probability of following a link rather than jumping to a random node.
    \item Betweenness Centrality (\(C_B(v)\)): Measures how often a node acts as a bridge in the shortest paths between other nodes.
    \begin{equation}
    C_B(v) = \sum_{s,t \in V_{n}^{C_i}, s \neq t} \frac{\sigma_{st}(v)}{\sigma_{st}}
    \end{equation}
    where \(\sigma_{st}\) is the total number of shortest paths from node \(s\) to node \(t\), and \(\sigma_{st}(v)\) is the number of those paths that pass through \(v\).
    \item Closeness Centrality (\(C_C(v)\)): Measures how efficiently a node can reach all other nodes in the community.
    \begin{equation}
    C_C(v) = \frac{|V_{n}^{C_i}| - 1}{\sum_{u \in V_{n}^{C_i}} d(v, u)}
    \end{equation}
    where \(d(v, u)\) is the shortest path distance between \(v\) and \(u\) in the community.
    \item Eigenvector Centrality (\(C_E(v)\)): Measures a node’s influence by considering the importance of its connected nodes.
    \begin{equation}
    C_E(v) = \frac{1}{\lambda} \sum_{u \in C_i} A_{vu} C_{E}(u)
    \end{equation}
    where \(\lambda\) is the largest eigenvalue of the adjacency matrix.
\end{itemize}

The final ranking score for each node \(v\) is computed as a weighted aggregation of its centrality values:
\begin{equation} 
R(v) = w_1 C_{out}(v) + w_2 C_{PR}(v) + w_3 C_B(v) + w_4 C_C(v) + w_5 C_E(v)
\end{equation}
where \( w_i \) are tunable weights that control the contribution of each centrality measure. The top-ranked nodes serve as additional contextual nodes, ensuring that the extracted subgraph retains structural and behavioral significance.

Nodes with low centrality scores outside the set of high-centrality neighbors within the same community as the key API nodes are removed, thereby reducing graph complexity while preserving the essential context of key API interactions.

\textbf{b) GSF vectorization.} The optimized ACGs are transformed into sequences using the Depth-First Search (DFS) algorithm. A DistilBERT-based model is then leveraged to extract high-level sequential features from these sequences, forming the GSF feature dataset. The model captures contextual dependencies between API calls, preserving both structural and semantic relationships within the execution flow. The final hidden layer generates a compact GSF feature vector that encapsulates key behavioral characteristics of the APK’s API call sequences. Formally, the GSF feature dataset is defined as: \(
\mathbf{GSF\_dataset} = \{ v_1^{GSF}, v_2^{GSF}, \dots, v_m^{GSF} \}
\) where \( v_i^{GSF} \in \mathbb{R}^{1 \times 128} \) denotes the GSF feature vector of the \( j \)-th API call sequence \( S_j \).

\subsection{Feature fusion}
After transforming each unimodal feature into a uniform vector representation, these vectors are fused through concatenation, self-attention, and cross-attention, effectively integrating multimodal data into a unified representation. To further refine the fusion process, gated fusion is introduced to dynamically control the contribution of each modality, allowing the model to enhance or suppress specific feature sets selectively. Additionally, dynamic weighted fusion adaptively adjusts the importance of each modality based on its relevance to the classification task, ensuring that the most informative features are emphasized. 

\subsubsection{Concatenation}
Concatenation is the simplest form of fusion, which directly combines the feature vectors from each modality along the feature dimension:
\begin{equation} 
v^{concat}_{fusion} = \text{Concat}(v^{TF}, v^{IF}, v^{GSF}) \in \mathbb{R}^{B \times d_{fusion}}
\end{equation}
The resulting representation \(v_{fusion}\) will have the shape \(\mathbb{R}^{B \times d_{fusion}}\), where \(B\) is the batch size, and \( d_{fusion} = d_{TF} + d_{IF} + d_{GSF}\) is the total dimensionality of the fused feature vector obtained by concatenating the feature vectors from each modality.

\subsubsection{Self-attention fusion}
Self-attention mechanisms \cite{vaswani2017attention} are employed to refine intra-modal representations for each modality before fusion. It enables each input vector to attend to itself by computing weighted combinations through the use of Query \( Q \), Key \( K \), and Value \( V \) vectors. Precisely, attention scores are calculated using the dot product between \( Q \) and \( K \), then scaled by \(\sqrt{d_k}\) to avoid large values that could lead to vanishing gradients in the softmax. This operation is formulated as:
\begin{equation} \text{Attention}(Q, K, V) = \text{softmax} \left( \frac{Q K^\top}{\sqrt{d_k}} \right) V \end{equation}
where \(d_k\) is the dimensionality of the key vectors. 
To capture diverse semantic patterns, we apply multi-head attention by projecting \( Q \), \( K \), and \( V \)  into multiple subspaces using trainable weights \(W_{Q_i}\), \(W_{K_i}\), \(W_{V_i}\) for each head \(i\):
\begin{equation}
\text{head}_i = \text{Attention}(Q_{m}W_{Q_i}, K_{m}W_{K_i}, V_{m}W_{V_i}) \end{equation}
The outputs of all heads are then concatenated and linearly projected using \(W_O\):
\begin{equation}
v_{attn}^{m} = \text{Concat}(\text{head}_1, \text{head}_2, \dots, \text{head}_h)W_O
\end{equation}
where \(h\) is the number of attention heads. 

In our setup, for each modality \( m \in \{TF, IF, GSF\} \), we set \( Q_{m} = K_{m} = V_{m} = v^{m} \), and each \( v_{m} \in \mathbb{R}^{B \times 1 \times d_{m}} \), representing a batch of input modality features with sequence length 1 and feature dimension \( d_m \). Applying self-attention to each modality separately, we obtain refined intra-modal representations:
\begin{equation} v_{attn}^{TF}, v_{attn}^{IF}, v_{attn}^{GSF} = \text{MultiHeadAttn}(Q_m, K_m, V_m) \end{equation}

After applying self-attention, the outputs are squeezed to remove the singleton sequence dimension, resulting in 
\(
v_{attn}^{m} \in \mathbb{R}^{B \times d_{m}}
\). These attended modality-specific features are then concatenated into a unified representation:
\begin{equation} v_{fusion}^{self-attn} = \text{Concat}(v_{attn}^{TF}, v_{attn}^{IF}, v_{attn}^{GSF}) \in \mathbb{R}^{B \times d_{fusion}} \end{equation}

\subsubsection{Cross-attention fusion}  
Unlike self-attention, cross-attention enables inter-modality feature interactions, thereby improving overall contextual understanding. For each modality \( m \in \{TF, IF, GSF\} \), we define the query, key, and value matrices for cross-attention:
\begin{equation}
Q_{m} = v^{m}, \quad K_{m} = \text{Concat}(v^{n}, v^{p}), \quad V_{m} = \text{Concat}(v^{n}, v^{p})
\end{equation}
where \( m, n, p \in \{TF, IF, GSF\}\) and \( m \neq n \neq p \). Each feature vector \( v_{m} \in \mathbb{R}^{B \times d_{m}} \) is reshaped to \(\mathbb{R}^{B \times 1 \times d_{m}}\) before applying attention. This reshaping ensures that the attention mechanism works on the appropriate input shape. 

Using multi-head cross-attention, we compute the inter-modal representations:
\begin{equation}
v_{cross}^{TF}, v_{cross}^{IF}, v_{cross}^{GSF} = \text{MultiHeadAttn}(Q_m, K_m, V_m)
\end{equation}
The output of the cross-attention layer is again squeezed along the sequence dimension, resulting in \(v^{m}_{cross} \in \mathbb{R}^{B \times d_{m}}\).

Finally, the fused representation is obtained by concatenating along the feature dimension:
\begin{equation} v_{\text{fusion}}^{\text{cross}} = \text{Concat}(v_{\text{cross}}^{TF},\ v_{\text{cross}}^{IF},\ v_{\text{cross}}^{GSF}) \in \mathbb{R}^{B \times d_{fusion}} \end{equation}

This fused representation incorporates inter-modal dependencies and is expected to be more expressive than simple concatenation or self-attention-based fusion.

\subsubsection{Gated fusion} 
Gated fusion is a dynamic feature integration strategy that enables the model to control the contribution of each modality adaptively. Instead of treating all modalities equally, this method assigns a learned gate to each modality's feature vector, allowing the network to emphasize or suppress specific modalities based on the input context. 

To dynamically control the contribution of each modality, we learn a scalar gate value for each modality via a sigmoid-activated linear transformation:
\begin{equation} g^{TF} = \sigma(W_{g}^{TF} v^{TF}), \quad g^{IF} = \sigma(W_{g}^{IF} v^{IF}), \quad g^{GSF} = \sigma(W_{g}^{GSF} v^{GSF}) \end{equation}
where \(W_{g}^{m} \in \mathbb{R}^{d_{m} \times 1}\) are trainable parameters and \(\sigma(.)\) denotes the sigmoid activation. These gate values \(g^{m} \in \mathbb{R}^{B \times 1}\) lie in the range \([0,1]\) and act as soft selectors that dynamically modulate the contribution of each modality. The gated features are computed by element-wise multiplication between the gate values and the corresponding modality feature vectors:
\begin{equation}
\tilde{v}^{TF} = g^{TF} \odot v^{TF}, \quad 
\tilde{v}^{IF} = g^{IF} \odot v^{IF}, \quad 
\tilde{v}^{GSF} = g^{GSF} \odot v^{GSF}
\end{equation}

Finally, these gated features are fused using element-wise summation:
\begin{equation} v_{\text{fusion}}^{\text{gated}} = \tilde{v}^{TF} + \tilde{v}^{IF} + \tilde{v}^{GSF} \in \mathbb{R}^{B \times d_{gated}} \end{equation}
where  \(d_{gated}\) denotes the unified feature dimension, and all modality-specific representations are projected to the same dimension, i.e., \(d_{gated} = d_{IF} = d_{TF} = d_{GSF}\), to enable element-wise fusion.

\subsubsection{Dynamic weighted fusion}
Dynamic weighted fusion (DWF) aims to combine features from multiple modalities by assigning adaptive weights to each modality during the fusion process. The weights are dynamically calculated based on the importance of each modality for the specific task. Given the feature representations \( v_m \) for each modality \( m \in \{TF, IF, GSF\} \), we apply modality-specific attention to learn the importance of each modality.

First, we compute modality-specific attention scores using the dot product between each modality’s feature representation and a learnable weight vector \( w_m \). The attention score for each modality is then computed as:
\begin{equation}
s_m = w_m^\top v_m
\end{equation}
where \( w_m \in \mathbb{R}^{d_m} \) is the learnable weight vector for modality \( m \), and \( v_m \in \mathbb{R}^{B \times d_m} \) is the feature vector of modality \( m \).

Once the attention scores \( \alpha_m \) are obtained, they are normalized using the softmax function to ensure that the weights sum to 1:

\begin{equation}
\alpha_m = \frac{e^{s_m}}{\sum_{m'} e^{s_{m'}}}
\end{equation}
where \( \sum_{m'} \) represents the sum over all modalities. This normalization ensures that the fusion weights are interpretable as probabilities.

After calculating the weights, the dynamic weighted fusion of the modalities is computed as the weighted sum of the individual modality features:
\begin{equation}
v_{fusion}^{DWF} = \sum_{m} \alpha_m v_m
\end{equation}

Finally, the fused representation \( v_{fusion}^{DWF} \in \mathbb{R}^{B \times d_{fusion}} \) is passed through a final linear transformation to project the features into a desired dimensionality \( d_{fusion} \):
\begin{equation}
v_{fusion}^{DWF} = \text{Concat}(\alpha_{TF} v_{TF}, \alpha_{IF} v_{IF}, \alpha_{GSF} v_{GSF}) W_{fusion}
\end{equation}
where \( W_{fusion} \) is a learnable weight matrix, and the concatenated vector is projected to the final fused representation \( v_{fusion}^{DWF} \).

\subsection{Classification}

The final fused representation \(v_{fusion}\) is passed through a classification head consisting of a fully connected layer to produce the logit output. Since the task is binary classification, the model is trained using the Binary Cross-Entropy with Logits Loss (BCEWithLogitsLoss). This loss function internally applies a sigmoid activation, making it suitable for learning from raw logits and numerically more stable than applying sigmoid followed by standard binary cross-entropy. This setup enables the model to learn discriminative representations suitable for distinguishing between benign and malicious samples.

\section{Experiments} \label{sec_experiment}

\subsection{Research questions}
We aim to answer the following research questions (RQ) through experiments and analyses.

\begin{itemize}
    \item RQ1: How does the multimodal malware detection model perform compared to unimodal and bimodal baselines under different fusion strategies?
    \item RQ2: How robust is the multimodal malware detector to obfuscated samples compared to unimodal and bimodal baselines under different fusion strategies?
    \item RQ3: How robust is the multimodal malware detector to AE attacks compared to unimodal and bimodal baselines under different fusion strategies?
    \item RQ4: What are the trade-offs between detection performance and computational cost among unimodal and multimodal models?
\end{itemize}

\subsection{Experimental setup}
\subsubsection{Implementation}
All stages of data collection, feature extraction, and data preprocessing are performed on a virtual machine running Ubuntu 20.04, equipped with 256 GB of RAM, a 48-core CPU, and a 500 GB hard disk. Models were trained and evaluated on Kaggle’s cloud-based GPU platform, which provides an NVIDIA Tesla P100 GPU and 16 GB of RAM. All models were implemented in Python, using PyTorch as the primary deep learning framework.

\subsubsection{Unimodal baselines} \label{sec:uni-base}
To investigate the effectiveness of multimodal fusion in malware detection, we first construct three unimodal baseline models, each trained on a single modality (TF, IF, or GSF). Their detailed architectures with experimentally optimized configurations are summarized in \textbf{Table \ref{tab:unimodal_models}}.

\begin{itemize}
    \item \textbf{U1}: \textit{an MLP trained solely on TF features.} The input dimension is set to 966 after applying PCA on the original 10,590 unique TF features from our dataset, as described in \textbf{Section \ref{sec:tab-rep}}. The model consists of three fully connected layers (FC) with batch normalization (BatchNorm1d) and dropout to improve generalization. The hidden layers map the input to 256, 256, and 128 dimensions, respectively, before projecting to the output layer.
    \item \textbf{U2}: \textit{a CNN trained on IF features.} The input is represented as $3 \times 64 \times 64$ RGB images derived from the IF feature set, enabling convolutional layers to capture spatial and structural patterns of DEX files. 
    The model includes two convolutional layers (Conv) followed by max-pooling (Pool), a flattening step (Flatten), and two fully connected layers that reduce the representation from 57,600 to 128 dimensions before classification.
    \item \textbf{U3}: \textit{a DistilBERT model trained on GSF features.} DistilBERT is a lightweight variant of BERT that preserves about 97\% of its accuracy while being 40\% more compact and running approximately 60\% faster, making it well-suited for environments with limited computational resources. It can process input sequences of up to 512 tokens. In this setup, each API sequence is tokenized and passed through the DistilBERT encoder (6 transformer layers, 12 attention heads, hidden size 768). The representation of the [CLS] token is then fed into a fully connected layer with 128 neurons and a dropout rate of 0.3, followed by the final classification layer.
\end{itemize}

For GSF features in particular, we further refine their representation by selecting the top 30 key APIs with the highest scores, as illustrated in \textbf{Figure \ref{fig:common-key-apis}}. For each key API, we retain the top 5 neighboring nodes with the highest centrality values within their respective communities. These values were chosen after extensive experimentation, which indicated that such dimensionality reduction preserves detection performance while reducing computational time and cost. In addition, we evaluated various pre-trained language models, including BERT, DistilBERT, CodeBERT, GPT-2, BigBird, and Longformer. Among them, DistilBERT was found to be the most lightweight while still achieving the best overall effectiveness.

\begin{table}[!ht]
\centering
\small
\caption{Architectures of unimodal baseline models.}
\label{tab:unimodal_models}
\begin{tabular}{|c|c|c|c|}
\hline
\textbf{Model}      & \textbf{Layer}    & \textbf{Input Dim} & \textbf{Out Dim}                                               \\ \hline
\multirow{5}{*}{U1} & FC1 + BatchNorm1d & 966                & 256                                                            \\
                    & FC2 + BatchNorm1d & 256                & 256                                                            \\
                    & Dropout           & 256                & 256                                                            \\
                    & FC3 + BatchNorm1d & 256                & 128                                                            \\
                    & FC4 (Output)      & 128                & 1                                                              \\ \hline
\multirow{6}{*}{U2} & Conv1             & (3, 64, 64)        & (32, 62, 62)                                                   \\
                    & Conv2             & (32, 62, 62)       & (64, 60, 60)                                                   \\
                    & Pool              & (64, 60, 60)       & (64, 30, 30)                                                   \\
                    & Flatten           & (64, 30, 30)       & 57.600                                                         \\
                    & FC1               & 57.600             & 128                                                            \\
                    & FC2 (Output)      & 128                & 1                                                              \\ \hline
\multirow{4}{*}{U3} & Encoder           & $N$ ($\leq$ 512)   & ($N$, 768)                                                     \\
                    & CLS token         & ($N$, 768)         & 768                                                            \\
                    & FC (Hidden)       & 768                & 128                                                            \\
                    & FC (Output)       & 128                & 1                                                              \\ \hline
\end{tabular}
\end{table}

\subsubsection{Multimodal configurations} \label{sec:multi-conf}

Our study focuses on intermediate fusion, where the high-level feature representations 
(128-dimensional outputs from each unimodal model) are fused to combine complementary 
information across modalities. Building on the unimodal baselines, we explore how different modalities complement each other by constructing five multimodal models that fuse TF, IF, and GSF features using different fusion strategies:

\begin{itemize}
    \item \textbf{M1}: Concatenation fusion.
    \item \textbf{M2}: Self-attention fusion.
    \item \textbf{M3}: Cross-attention fusion.
    \item \textbf{M4}: Gated fusion.
    \item \textbf{M5}: Dynamic weighted fusion.
\end{itemize}

To provide a comprehensive understanding of how each modality contributes to the multimodal setting, we further constructed 15 bimodal models by pairing two out of the three modalities and applying the same five fusion strategies as in the multimodal setting. These bimodal configurations serve as intermediate baselines to analyze the impact and complementarity of each modality when combined, before fully integrating all three in the multimodal models.

\subsubsection{Training and evaluation details}
The training process employed the \(AdamW\) optimizer with a learning rate of \(2e-5\) and a weight decay of \(0.01\) to enhance generalization and prevent overfitting. A batch size of \(32\) is utilized to balance computational efficiency and model convergence. The model is trained for \(30\) epochs to ensure sufficient learning while mitigating the risk of overfitting. The Binary Cross Entropy with Logits (BCEWithLogitsLoss) function is used as the loss criterion, effectively handling binary classification tasks by combining a sigmoid activation with binary cross-entropy loss.

\subsubsection{Dataset and metrics} 
To evaluate our proposed approach, we utilize a curated dataset comprising 16,691 Android applications, including 4,030 benign samples and 12,661 malware samples, sourced from the CICMalDroid 2020 dataset \cite{mahdavifar2022effective}. This dataset collects samples from multiple reputable sources, including VirusTotal, Contagio security blog, AMD, and MalDozer, encompassing five distinct malware categories, including Adware, Banking malware, Riskware, and SMS malware. All samples undergo an initial dynamic analysis phase using CopperDroid, a virtual machine introspection (VMI)-based analysis framework. To maintain data integrity, samples that fail during execution due to time-outs, invalid APK files, and memory allocation errors are excluded from further analysis.

The dataset is split into 70\% for training and 30\% for testing. To further evaluate the robustness and resilience of our multimodal fusion approach, the test set is also employed to generate obfuscated and adversarial samples, as discussed in \textbf{Sections \ref{Code_obfuscation}} and \textbf{\ref{adversarial_example}}.

We employ Accuracy, Recall, Precision, and F1-score as the primary metrics for model evaluation. Accuracy measures the overall proportion of correct predictions, providing a general sense of model performance. Recall quantifies the ability of the model to correctly identify benign applications among all actual benign samples, while Precision assesses the correctness of those identified as benign. The F1-score, as the harmonic mean of precision and recall, offers a balanced view of the model’s effectiveness, particularly in the presence of class imbalance. The detailed results for these metrics are presented in \textbf{Table \ref{tab:metrics}}.

\begin{table}[!ht]
\centering
\small
\caption{Evaluation metrics.}
\begin{tabular}{|l|l|}
\hline
\textbf{Metrics}    & \textbf{Description}                     \\ \hline
True Positive (TP)  & Malware correctly classified as malware  \\ \hline
True Negative (TN)  & Benign correctly classified as benign    \\ \hline
False Positive (FP) & Benign incorrectly classified as malware \\ \hline
False Negative (FN) & Malware incorrectly classified as benign \\ \hline
Accuracy (Acc)      & $(TP+TN)/(TP+TN+FP+FN)$                  \\ \hline
Precision (Pre)     & $TP/(TP+FP)$                             \\ \hline
Recall (Rec)        & $TP/(TP+FN)$                             \\ \hline
F1-score (F1)       & $2 \cdot (\text{Pre} \cdot \text{Rec})/(\text{Pre} + \text{Rec})$ \\ \hline
\end{tabular}
\label{tab:metrics}
\end{table}

\subsubsection{Code obfuscation}
\label{Code_obfuscation}
To evaluate model resilience in \textbf{RQ2}, we employ Obfuscapk \cite{aonzo2020obfuscapk}, a state-of-the-art open-source black-box obfuscation tool that takes an Android application as input and outputs a new one with its bytecode obfuscated. We selected a representative set of obfuscation strategies that are widely used in the wild to conceal malicious behaviors \cite{molina2023towards}. These strategies are grouped into three main categories, along with a fourth mixed scenario to simulate layered attacks.

\begin{itemize}
    \item \textbf{Renaming (Rn):} Replaces meaningful identifiers (classes, methods, fields) with obscure ones to strip semantic clues. Class or package renaming requires updating the AndroidManifest.xml, making the app appear as a new package in the ecosystem.

    \item \textbf{Code obfuscation (Co):} Applies bytecode-level transformations such as \textit{CallIndirection}, \textit{Goto}, \textit{Reorder}, \textit{ArithmeticBranch}, \textit{Nop}, and \textit{MethodsOverload}. These techniques distort control flow and insert junk instructions, complicating reverse engineering and static analysis.

    \item \textbf{Encryption (Enc):} Native libraries, strings, and assets are encrypted with a randomly generated key and decrypted only at runtime. This conceals sensitive resources from static analysis but adds performance overhead due to repeated decryption.

    \item \textbf{Mixed obfuscation (Rn+Co+Enc):} Combines all techniques to emulate sophisticated real-world malware that layers multiple obfuscation strategies, producing the most challenging obfuscated samples for evaluation. Prior studies have shown that while code-structure modifications and encryption are among the most effective standalone strategies, their combination with other techniques further amplifies evasion capability \cite{molina2023towards}. In fact, most leading commercial anti-malware tools fail to detect malicious applications obfuscated with multiple strategies applied simultaneously \cite{nawaz2022evaluation}.
\end{itemize}

\subsubsection{Adversarial examples}
\label{adversarial_example}
In \textbf{RQ3}, we evaluate model robustness against AEs generated from malware samples in the test set. Currently, there is an increasing demand for AEs that remain realistic and executable, meaning they must preserve the malicious behaviors of the original samples \cite{bae2021learn2evade}. Unlike images, where perturbations such as changing a pixel value do not compromise the object’s semantics, even small modifications in malware (e.g., altering or removing API calls) can easily disrupt its functionality. While directly modifying functional features such as API calls is highly disruptive and difficult to scale, permissions and intents provide a more feasible alternative. These features can be changed without affecting malicious behavior, and malware typically requests more sensitive ones than benign apps, making them strong indicators for detection. Building on this insight, and inspired by MalGAN \cite{hu2022generating}, we redesign this GAN-based framework to generate more realistic adversarial malware samples. Specifically, we create AEs by randomly adding or removing only permissions and intents frequently associated with benign applications, while preserving the malicious behavioral features.

Notably, the generated AEs also demonstrate strong transferability across different detection models. In our setup, the discriminator is implemented as a DL model pretrained on a dataset labeled by the black-box baseline model U1 described in \textbf{Section \ref{sec:uni-base}}. This design enables the generator to learn perturbations that remain effective even when transferred to unseen detectors. We assume that malware authors only know the type of features used by this black-box detector but have no knowledge of the underlying learning algorithm or access to model parameters. In practice, they can only query the detector to obtain binary detection results for their programs. 

To ensure behavioral integrity, each APK is installed and executed in an Android Emulator using the Monkey tool. We compare the behavior logs of the original and adversarial apps to verify semantic consistency, following the validation methodology described in \cite{renjith2022gang}. After this validation, over 95\% of the AEs remain functional and preserve their original malicious behavior.

\subsection{Experimental results}

\subsubsection{RQ1}
\textbf{Table \ref{tab:ori-comparison}} presents the experimental results of unimodal and multimodal configurations. Among unimodal baselines, U1 achieves the best performance, with 96.79\% accuracy and 97.88\% F1, demonstrating its robustness. U2 also performs competitively, achieving 94.65\% accuracy and 96.50\% F1, but with a slightly lower recall of 97.18\% compared to U1. By contrast, U3 records the lowest accuracy at 87.46\%, yet maintains a relatively strong recall of 92.47\%, suggesting that sequence-based features are particularly effective at capturing malicious behaviors even though they sacrifice precision and overall accuracy.

In the multimodal setting, all fusion strategies outperform their unimodal counterparts, confirming the benefit of combining complementary modalities. Although U3 performs worse as a standalone model, its integration with other modalities consistently improves overall performance. For example, in the bimodal case, U1+U3-DWF achieves 97.32\% accuracy and 98.24\% F1, while U2+U3-DWF improves to 95.87\% accuracy and 97.28\% F1, both surpassing their respective unimodal baselines. This demonstrates that API sequences provide complementary information that enhances detection performance, with DWF particularly effective in balancing the contributions of different modalities compared to other fusion mechanisms.

The advantage becomes even more pronounced when all three modalities are fused. Across all fusion strategies (M1 to M5), the trimodal models achieve better performance than most unimodal and bimodal baselines. Notably, M3 with cross-attention fusion and M5 with DWF achieve robust results, with M3 reaching 98.04\% accuracy and 98.71\% F1, while M5 achieves 97.98\% accuracy and 98.67\% F1.

The only exception is U1+U2-DWF, which performs slightly better than the trimodal models. However, such marginal gains raise the question of whether this bimodal configuration can maintain robustness under adversarial attacks and code obfuscation, where trimodal fusion may offer greater stability.

\begin{table}[!ht]
\centering
\small
\caption{Unimodal and multimodal malware detection performance comparison under original dataset (\%).}
\label{tab:ori-comparison}
\begin{tabular}{|c|c|c|c|c|}
\hline
\textbf{Modality} & \textbf{Accuracy}             & \textbf{Recall}               & \textbf{Precision}            & \textbf{F1-score}             \\ \hline
U1                & 96.79                         & 97.89                         & 97.87                         & 97.88                         \\ \hline
U2                & 94.65                         & 97.18                         & 95.82                         & 96.50                         \\ \hline
U3                & 87.46                         & 92.47                         & 91.13                         & 91.80                         \\ \hline
U1+U2-concat    & 98.10                         & 99.37                         & 98.15                         & 98.76                         \\ \hline
U1+U3-concat    & 97.36                         & 98.13                         & 98.39                         & 98.26                         \\ \hline
U2+U3-concat    & 93.13                         & 95.68                         & 95.28                         & 95.48                         \\ \hline
U1+U2-attn      & 98.36                         & 99.47                         & 98.39                         & 98.93                         \\ \hline
U1+U3-attn      & 97.34                         & 98.42                         & 98.08                         & 98.25                         \\ \hline
U2+U3-attn      & 93.83                         & 95.60                         & 96.24                         & 95.92                         \\ \hline
U1+U2-cross     & 98.06                         & 99.53                         & 97.95                         & 98.73                         \\ \hline
U1+U3-cross     & 96.87                         & 98.03                         & 97.85                         & 97.94                         \\ \hline
U2+U3-cross     & 94.57                         & 96.10                         & 96.72                         & 96.41                         \\ \hline
U1+U2-gated     & 98.20                         & 99.34                         & 98.31                         & 98.82                         \\ \hline
U1+U3-gated     & 97.02                         & 98.82                         & 97.30                         & 98.05                         \\ \hline
U2+U3-gated     & 93.37                         & 94.37                         & 96.81                         & 95.57                         \\ \hline
\textbf{U1+U2-DWF}       & \textbf{98.54}                         & \textbf{98.95}                         & \textbf{99.13}                         & \textbf{99.04}                         \\ \hline
U1+U3-DWF       & 97.32                         & 98.32                         & 98.16                         & 98.24                         \\ \hline
U2+U3-DWF       & 95.87                         & 97.45                         & 97.11                         & 97.28                         \\ \hline
M1                & 97.72 & 98.58 & 98.42 & 98.50 \\ \hline
M2                & 97.94 & 98.97 & 98.02 & 98.49 \\ \hline
\textbf{M3}                & \textbf{98.04} & \textbf{98.89} & \textbf{98.53} & \textbf{98.71} \\ \hline
M4                & 97.44 & 98.13 & 98.49 & 98.31 \\ \hline
\textbf{M5}                & \textbf{97.98} & \textbf{98.32} & \textbf{99.02} & \textbf{98.67} \\ \hline
\end{tabular}
\end{table}

\subsubsection{RQ2}
\textbf{Table \ref{tab:obf-compare-no-full}} shows the performance of unimodal and multimodal models under different obfuscation scenarios, including renaming, code obfuscation, encryption, and their combination. Overall, the results show that multimodal detectors exhibit substantially greater robustness compared to unimodal baselines. 

Among unimodal models, U1 remains relatively stable across obfuscation techniques, with only marginal drops in recall, from 97.77\% under renaming to 97.68\% under Rn+Co+Enc. This stability arises because permissions and intents are predefined by the Android framework and applications, with core permissions fixed by the system and thus unaffected by obfuscation. By contrast, U2 is more sensitive to code obfuscation, with performance dropping to 89.76\% recall and 92.38\% F1. However, when combined with renaming and encryption, the impact is less severe, as renaming does not affect the underlying bytecode structure, while encryption may introduce distinctive global patterns (e.g., decryption routines, unusual padding, higher entropy) that potentially help CNN distinguish malware from benign apps. Meanwhile, U3 remains largely unaffected, since obfuscation cannot alter system-defined API calls. Consequently, DistilBERT still maintains or even improves recall and F1.

Since obfuscation does not severely impact unimodal models, multimodal models not only maintain performance but also substantially boost accuracy and F1, achieving overall improvements of around 2–5\% compared to the best unimodal baselines. Among different fusion strategies, M5 consistently delivers the highest scores, with recall, precision, and F1 values reaching up to 99\% across all obfuscation scenarios. In general, these results demonstrate that leveraging all three information sources, including permissions and intents, bytecode, and API sequences, makes the system highly resilient and very difficult to evade through obfuscation.

\begin{table*}[!ht]
\centering
\small
\caption{Performance of unimodal and multimodal models under different obfuscation techniques.}
\label{tab:obf-compare-no-full}
\scalebox{0.9}{\begin{tabular}{|c|cccc|cccc|cccc|cccc|}
\hline
\multirow{2}{*}{\textbf{Modality}} & \multicolumn{4}{c|}{\textbf{Renaming (Rn)}}                                                                                      & \multicolumn{4}{c|}{\textbf{Code obfuscation (Co)}}                                                                              & \multicolumn{4}{c|}{\textbf{Encryption (Enc)}}                                                                                   & \multicolumn{4}{c|}{\textbf{Rn+Co+Enc}}                                                                                          \\ \cline{2-17} 
                                   & \multicolumn{1}{c|}{\textbf{Acc}}   & \multicolumn{1}{c|}{\textbf{Rec}}   & \multicolumn{1}{c|}{\textbf{Pre}}   & \textbf{F1}    & \multicolumn{1}{c|}{\textbf{Acc}}   & \multicolumn{1}{c|}{\textbf{Rec}}   & \multicolumn{1}{c|}{\textbf{Pre}}   & \textbf{F1}    & \multicolumn{1}{c|}{\textbf{Acc}}   & \multicolumn{1}{c|}{\textbf{Rec}}   & \multicolumn{1}{c|}{\textbf{Pre}}   & \textbf{F1}    & \multicolumn{1}{c|}{\textbf{Acc}}   & \multicolumn{1}{c|}{\textbf{Rec}}   & \multicolumn{1}{c|}{\textbf{Pre}}   & \textbf{F1}    \\ \hline
U1                                 & \multicolumn{1}{c|}{96.58}          & \multicolumn{1}{c|}{97.77}          & \multicolumn{1}{c|}{97.59}          & 97.68          & \multicolumn{1}{c|}{96.83}          & \multicolumn{1}{c|}{98.04}          & \multicolumn{1}{c|}{97.71}          & 97.88          & \multicolumn{1}{c|}{96.85}          & \multicolumn{1}{c|}{98.07}          & \multicolumn{1}{c|}{97.71}          & 97.89          & \multicolumn{1}{c|}{96.43}          & \multicolumn{1}{c|}{97.68}          & \multicolumn{1}{c|}{97.36}          & 97.52          \\ \hline
U2                                 & \multicolumn{1}{c|}{95.23}          & \multicolumn{1}{c|}{98.30}          & \multicolumn{1}{c|}{95.35}          & 96.80          & \multicolumn{1}{c|}{\textbf{88.98}}          & \multicolumn{1}{c|}{\textbf{89.76}}          & \multicolumn{1}{c|}{\textbf{95.16}}          & \textbf{92.38}          & \multicolumn{1}{c|}{95.27}          & \multicolumn{1}{c|}{98.21}          & \multicolumn{1}{c|}{95.56}          & 96.87          & \multicolumn{1}{c|}{94.96}          & \multicolumn{1}{c|}{98.23}          & \multicolumn{1}{c|}{94.91}          & 96.54          \\ \hline
U3                                 & \multicolumn{1}{c|}{92.07}          & \multicolumn{1}{c|}{99.40}          & \multicolumn{1}{c|}{90.71}          & 94.86          & \multicolumn{1}{c|}{91.62}          & \multicolumn{1}{c|}{98.41}          & \multicolumn{1}{c|}{91.05}          & 94.59          & \multicolumn{1}{c|}{91.61}          & \multicolumn{1}{c|}{98.44}          & \multicolumn{1}{c|}{91.03}          & 94.59          & \multicolumn{1}{c|}{88.39}          & \multicolumn{1}{c|}{94.99}          & \multicolumn{1}{c|}{89.45}          & 92.14          \\ \hline
U1+U2-concat                       & \multicolumn{1}{c|}{97.81}          & \multicolumn{1}{c|}{99.14}          & \multicolumn{1}{c|}{97.91}          & 98.52          & \multicolumn{1}{c|}{96.47}          & \multicolumn{1}{c|}{97.28}          & \multicolumn{1}{c|}{97.97}          & 97.62          & \multicolumn{1}{c|}{98.12}          & \multicolumn{1}{c|}{99.49}          & \multicolumn{1}{c|}{98.02}          & 98.75          & \multicolumn{1}{c|}{97.77}          & \multicolumn{1}{c|}{99.21}          & \multicolumn{1}{c|}{97.71}          & 98.46          \\ \hline
U1+U3-concat                       & \multicolumn{1}{c|}{97.88}          & \multicolumn{1}{c|}{98.93}          & \multicolumn{1}{c|}{98.20}          & 98.56          & \multicolumn{1}{c|}{98.44}          & \multicolumn{1}{c|}{99.63}          & \multicolumn{1}{c|}{98.29}          & 98.96          & \multicolumn{1}{c|}{98.44}          & \multicolumn{1}{c|}{99.63}          & \multicolumn{1}{c|}{98.29}          & 98.96          & \multicolumn{1}{c|}{97.58}          & \multicolumn{1}{c|}{98.62}          & \multicolumn{1}{c|}{98.01}          & 98.32          \\ \hline
U2+U3-concat                       & \multicolumn{1}{c|}{95.42}          & \multicolumn{1}{c|}{99.14}          & \multicolumn{1}{c|}{94.87}          & 96.96          & \multicolumn{1}{c|}{95.16}          & \multicolumn{1}{c|}{98.58}          & \multicolumn{1}{c|}{95.10}          & 96.81          & \multicolumn{1}{c|}{95.35}          & \multicolumn{1}{c|}{98.87}          & \multicolumn{1}{c|}{95.09}          & 96.94          & \multicolumn{1}{c|}{94.39}          & \multicolumn{1}{c|}{98.07}          & \multicolumn{1}{c|}{94.33}          & 96.16          \\ \hline
U1+U2-attn                         & \multicolumn{1}{c|}{98.12}          & \multicolumn{1}{c|}{99.29}          & \multicolumn{1}{c|}{98.17}          & 98.73          & \multicolumn{1}{c|}{96.92}          & \multicolumn{1}{c|}{97.62}          & \multicolumn{1}{c|}{98.23}          & 97.92          & \multicolumn{1}{c|}{98.42}          & \multicolumn{1}{c|}{99.63}          & \multicolumn{1}{c|}{98.27}          & 98.94          & \multicolumn{1}{c|}{98.08}          & \multicolumn{1}{c|}{99.35}          & \multicolumn{1}{c|}{98.00}          & 98.67          \\ \hline
U1+U3-attn                         & \multicolumn{1}{c|}{97.33}          & \multicolumn{1}{c|}{98.54}          & \multicolumn{1}{c|}{97.84}          & 98.19          & \multicolumn{1}{c|}{98.29}          & \multicolumn{1}{c|}{99.77}          & \multicolumn{1}{c|}{97.97}          & 98.86          & \multicolumn{1}{c|}{98.29}          & \multicolumn{1}{c|}{99.77}          & \multicolumn{1}{c|}{97.97}          & 98.86          & \multicolumn{1}{c|}{97.51}          & \multicolumn{1}{c|}{98.92}          & \multicolumn{1}{c|}{97.64}          & 98.28          \\ \hline
U2+U3-attn                         & \multicolumn{1}{c|}{96.58}          & \multicolumn{1}{c|}{99.58}          & \multicolumn{1}{c|}{95.93}          & 97.72          & \multicolumn{1}{c|}{96.16}          & \multicolumn{1}{c|}{98.84}          & \multicolumn{1}{c|}{96.11}          & 97.46          & \multicolumn{1}{c|}{96.16}          & \multicolumn{1}{c|}{98.87}          & \multicolumn{1}{c|}{96.08}          & 97.46          & \multicolumn{1}{c|}{95.57}          & \multicolumn{1}{c|}{98.46}          & \multicolumn{1}{c|}{95.49}          & 96.95          \\ \hline
U1+U2-cross                        & \multicolumn{1}{c|}{98.25}          & \multicolumn{1}{c|}{99.73}          & \multicolumn{1}{c|}{97.92}          & 98.82          & \multicolumn{1}{c|}{96.87}          & \multicolumn{1}{c|}{97.82}          & \multicolumn{1}{c|}{97.98}          & 97.90          & \multicolumn{1}{c|}{98.37}          & \multicolumn{1}{c|}{99.83}          & \multicolumn{1}{c|}{98.02}          & 98.92          & \multicolumn{1}{c|}{98.12}          & \multicolumn{1}{c|}{99.71}          & \multicolumn{1}{c|}{97.72}          & 98.70          \\ \hline
U1+U3-cross                        & \multicolumn{1}{c|}{96.96}          & \multicolumn{1}{c|}{97.44}          & \multicolumn{1}{c|}{98.41}          & 97.92          & \multicolumn{1}{c|}{97.89}          & \multicolumn{1}{c|}{98.67}          & \multicolumn{1}{c|}{98.50}          & 98.58          & \multicolumn{1}{c|}{97.89}          & \multicolumn{1}{c|}{98.67}          & \multicolumn{1}{c|}{98.50}          & 98.58          & \multicolumn{1}{c|}{97.02}          & \multicolumn{1}{c|}{97.58}          & \multicolumn{1}{c|}{98.25}          & 97.91          \\ \hline
U2+U3-cross                        & \multicolumn{1}{c|}{96.01}          & \multicolumn{1}{c|}{99.70}          & \multicolumn{1}{c|}{95.11}          & 97.35          & \multicolumn{1}{c|}{95.52}          & \multicolumn{1}{c|}{98.87}          & \multicolumn{1}{c|}{95.30}          & 97.05          & \multicolumn{1}{c|}{95.61}          & \multicolumn{1}{c|}{98.98}          & \multicolumn{1}{c|}{95.30}          & 97.11          & \multicolumn{1}{c|}{95.1}           & \multicolumn{1}{c|}{98.79}          & \multicolumn{1}{c|}{94.61}          & 96.65          \\ \hline
U1+U2-gated                        & \multicolumn{1}{c|}{98.12}          & \multicolumn{1}{c|}{99.37}          & \multicolumn{1}{c|}{98.09}          & 98.73          & \multicolumn{1}{c|}{96.75}          & \multicolumn{1}{c|}{97.48}          & \multicolumn{1}{c|}{98.14}          & 97.81          & \multicolumn{1}{c|}{98.37}          & \multicolumn{1}{c|}{99.66}          & \multicolumn{1}{c|}{98.18}          & 98.92          & \multicolumn{1}{c|}{98.01}          & \multicolumn{1}{c|}{99.35}          & \multicolumn{1}{c|}{97.90}          & 98.62          \\ \hline
U1+U3-gated                        & \multicolumn{1}{c|}{97.28}          & \multicolumn{1}{c|}{99.40}          & \multicolumn{1}{c|}{96.98}          & 98.18          & \multicolumn{1}{c|}{97.63}          & \multicolumn{1}{c|}{99.74}          & \multicolumn{1}{c|}{97.16}          & 98.43          & \multicolumn{1}{c|}{97.61}          & \multicolumn{1}{c|}{99.74}          & \multicolumn{1}{c|}{97.13}          & 98.42          & \multicolumn{1}{c|}{96.90}          & \multicolumn{1}{c|}{99.08}          & \multicolumn{1}{c|}{96.68}          & 97.87          \\ \hline
U2+U3-gated                        & \multicolumn{1}{c|}{95.99}          & \multicolumn{1}{c|}{99.70}          & \multicolumn{1}{c|}{95.11}          & 97.35          & \multicolumn{1}{c|}{95.52}          & \multicolumn{1}{c|}{98.87}          & \multicolumn{1}{c|}{95.30}          & 97.05          & \multicolumn{1}{c|}{95.82}          & \multicolumn{1}{c|}{98.98}          & \multicolumn{1}{c|}{95.30}          & 97.11          & \multicolumn{1}{c|}{94.09}          & \multicolumn{1}{c|}{98.79}          & \multicolumn{1}{c|}{94.61}          & 96.65          \\ \hline
U1+U2-DWF                          & \multicolumn{1}{c|}{98.95}          & \multicolumn{1}{c|}{99.55}          & \multicolumn{1}{c|}{99.02}          & 99.29          & \multicolumn{1}{c|}{97.09}          & \multicolumn{1}{c|}{97.02}          & \multicolumn{1}{c|}{99.04}          & 98.02          & \multicolumn{1}{c|}{99.07}          & \multicolumn{1}{c|}{99.69}          & \multicolumn{1}{c|}{99.07}          & 99.38          & \multicolumn{1}{c|}{98.92}          & \multicolumn{1}{c|}{99.57}          & \multicolumn{1}{c|}{98.93}          & 99.25          \\ \hline
U1+U3-DWF                          & \multicolumn{1}{c|}{97.66}          & \multicolumn{1}{c|}{98.90}          & \multicolumn{1}{c|}{97.94}          & 98.41          & \multicolumn{1}{c|}{97.76}          & \multicolumn{1}{c|}{98.98}          & \multicolumn{1}{c|}{98.03}          & 98.50          & \multicolumn{1}{c|}{97.76}          & \multicolumn{1}{c|}{98.98}          & \multicolumn{1}{c|}{98.03}          & 98.50          & \multicolumn{1}{c|}{97.42}          & \multicolumn{1}{c|}{98.69}          & \multicolumn{1}{c|}{97.73}          & 98.21          \\ \hline
U2+U3-DWF                          & \multicolumn{1}{c|}{97.26}          & \multicolumn{1}{c|}{99.55}          & \multicolumn{1}{c|}{96.81}          & 98.16          & \multicolumn{1}{c|}{96.62}          & \multicolumn{1}{c|}{98.58}          & \multicolumn{1}{c|}{96.93}          & 97.75          & \multicolumn{1}{c|}{96.96}          & \multicolumn{1}{c|}{99.04}          & \multicolumn{1}{c|}{96.95}          & 97.98          & \multicolumn{1}{c|}{96.58}          & \multicolumn{1}{c|}{98.82}          & \multicolumn{1}{c|}{96.48}          & 97.64          \\ \hline
M1                                 & \multicolumn{1}{c|}{98.58}          & \multicolumn{1}{c|}{99.85}          & \multicolumn{1}{c|}{98.24}          & 99.04          & \multicolumn{1}{c|}{98.31}          & \multicolumn{1}{c|}{99.43}          & \multicolumn{1}{c|}{98.32}          & 98.87          & \multicolumn{1}{c|}{98.52}          & \multicolumn{1}{c|}{99.72}          & \multicolumn{1}{c|}{98.32}          & 99.01          & \multicolumn{1}{c|}{98.08}          & \multicolumn{1}{c|}{98.67}          & \multicolumn{1}{c|}{99.28}          & 98.67          \\ \hline
M2                                 & \multicolumn{1}{c|}{98.69}          & \multicolumn{1}{c|}{99.67}          & \multicolumn{1}{c|}{98.56}          & 99.11          & \multicolumn{1}{c|}{98.69}          & \multicolumn{1}{c|}{99.63}          & \multicolumn{1}{c|}{98.62}          & 99.13          & \multicolumn{1}{c|}{98.82}          & \multicolumn{1}{c|}{99.80}          & \multicolumn{1}{c|}{98.63}          & 99.21          & \multicolumn{1}{c|}{98.26}          & \multicolumn{1}{c|}{99.18}          & \multicolumn{1}{c|}{98.41}          & 98.79          \\ \hline
M3                                 & \multicolumn{1}{c|}{97.88}          & \multicolumn{1}{c|}{99.85}          & \multicolumn{1}{c|}{97.33}          & 98.57          & \multicolumn{1}{c|}{97.87}          & \multicolumn{1}{c|}{99.72}          & \multicolumn{1}{c|}{97.48}          & 98.58          & \multicolumn{1}{c|}{97.85}          & \multicolumn{1}{c|}{99.72}          & \multicolumn{1}{c|}{97.45}          & 98.57          & \multicolumn{1}{c|}{97.72}          & \multicolumn{1}{c|}{99.84}          & \multicolumn{1}{c|}{97.07}          & 98.43          \\ \hline
M4                                 & \multicolumn{1}{c|}{98.62}          & \multicolumn{1}{c|}{99.82}          & \multicolumn{1}{c|}{98.33}          & 99.07          & \multicolumn{1}{c|}{98.54}          & \multicolumn{1}{c|}{99.66}          & \multicolumn{1}{c|}{98.40}          & 99.03          & \multicolumn{1}{c|}{98.65}          & \multicolumn{1}{c|}{99.80}          & \multicolumn{1}{c|}{98.41}          & 99.10          & \multicolumn{1}{c|}{98.17}          & \multicolumn{1}{c|}{99.31}          & \multicolumn{1}{c|}{98.16}          & 98.73          \\ \hline
\textbf{M5}                        & \multicolumn{1}{c|}{\textbf{98.93}} & \multicolumn{1}{c|}{\textbf{99.64}} & \multicolumn{1}{c|}{\textbf{98.91}} & \textbf{99.27} & \multicolumn{1}{c|}{\textbf{98.94}} & \multicolumn{1}{c|}{\textbf{99.63}} & \multicolumn{1}{c|}{\textbf{98.96}} & \textbf{99.29} & \multicolumn{1}{c|}{\textbf{99.01}} & \multicolumn{1}{c|}{\textbf{99.72}} & \multicolumn{1}{c|}{\textbf{98.96}} & \textbf{99.34} & \multicolumn{1}{c|}{\textbf{98.50}} & \multicolumn{1}{c|}{\textbf{99.12}} & \multicolumn{1}{c|}{\textbf{98.79}} & \textbf{98.95} \\ \hline
\end{tabular}}
\end{table*}

\subsubsection{RQ3} \label{adv-exp}

In this part, we evaluate the robustness of all models under AE attacks. As described in \textbf{Section \ref{adversarial_example}}, the AEs are generated with the explicit objective of targeting U1 by perturbing permission and intent features. Consequently, U1 suffers the most severe degradation, with accuracy dropping to 26.80\% and recall falling to only 4.65\%, resulting in an overall F1-score of just 8.71\%. This sharp decline confirms that adversarial perturbations crafted on manifest-level features can effectively mislead detectors that rely solely on this modality. By contrast, U2 and U3 remain largely unaffected because the attack does not modify their primary feature sources.

When examining bimodal detectors, results vary sharply depending on whether U1 is included. Models combining U1 and U2 perform the worst across all fusion strategies. In particular, U1+U2-DWF almost collapses with only 25.58\% accuracy and 1.85\% F1, showing that dynamic weighting cannot compensate for the severe vulnerability of U1 when paired with U2. Other U1-based pairs, such as U1+U3, also experience substantial degradation. Specifically, U1+U3-concat achieves 77.08\% F1, U1+U3-attn only 25.20\% F1, and U1+U3-cross 42.90\% F1. Among these, U1+U3-gated proves the most stable, reaching 84.96\% accuracy and 89.20\% F1, although the performance remains below that on the original test set, but still noticeably stronger than other U1-involved fusions.

A similar trend is observed in trimodal models. M1 and M2, which rely on concatenation and attention, inherit U1’s vulnerability and degrade substantially (76.51\% and 45.51\% F1, respectively). In contrast, M3 and M4 perform much better with 90.66\% and 93.07\% F1. Finally, M5 with DWF clearly outperforms all unimodal and bimodal baselines, delivering the strongest defense against AEs with 98.88\% accuracy and 99.26\% F1, far exceeding any other fusion strategy. 

Interestingly, this stands in stark contrast to the bimodal case, where only U1 and U2 are fused, resulting in DWF collapsing with 25.58\% accuracy and 1.85\% F1. The reason lies in how the softmax-based weighting operates. Since the weights must sum to one, a compromised modality such as U1 cannot be completely ignored, and its residual contribution continues to degrade the bimodal detector’s performance. However, in the trimodal settings, the presence of U3 introduces an additional reliable source of information. This allows the softmax mechanism to shift most of the confidence toward CNN and DistilBERT while assigning only minimal weight to U1. As a result, the DWF strategy can effectively suppress the negative impact of U1 and rely more heavily on the stable modalities.

\begin{table}[!ht]
\centering
\small
\begin{tabular}{|c|c|c|c|c|}
\hline
\textbf{Modality} & \textbf{Accuracy} & \textbf{Recall} & \textbf{Precision} & \textbf{F1-score} \\ \hline
\textbf{U1}                & \textbf{26.80}& \textbf{4.65}& \textbf{67.6}& \textbf{8.71}\\ \hline
U2                & 95.12& 97.93& 95.67& 96.79\\ \hline
U3                & 89.79& 95.81& 91.05& 93.37\\ \hline
U1+U2-concat    & 45.10& 28.78& 93.64& 44.03\\ \hline
U1+U3-concat    & 71.55& 63.76& 97.43& 77.08\\ \hline
U2+U3-concat    & 95.23& 98.60& 95.21& 96.87\\ \hline
U1+U2-attn      & 26.61& 3.88& 69.46& 7.36\\ \hline
U1+U3-attn      & 34.50& 14.71& 87.97& 25.20\\ \hline
U2+U3-attn      & 96.14& 98.76& 96.19& 97.46\\ \hline
U1+U2-cross     & 37.07& 18.29& 89.37& 30.36\\ \hline
U1+U3-cross     & 44.24& 27.93& 92.52& 42.90\\ \hline
U2+U3-cross     & 96.34& 98.54& 96.65& 97.59\\ \hline
U1+U2-gated     & 52.85& 38.94& 95.61& 55.34\\ \hline
U1+U3-gated     & 84.96& 82.81& 96.66& 89.20\\ \hline
U2+U3-gated     & 96.22& 98.54& 96.65& 97.59\\ \hline
\textbf{U1+U2-DWF}       & \textbf{25.58}& \textbf{0.94}& \textbf{87.18}& \textbf{1.85}\\ \hline
U1+U3-DWF       & 30.21& 7.33& 95.34& 13.61\\ \hline
U2+U3-DWF       & 96.86& 98.84& 97.03& 97.93\\ \hline
M1                & 70.99& 62.99& 97.44& 76.51\\ \hline
M2                & 46.36& 29.85& 95.68& 45.51\\ \hline
M3                & 86.98& 84.19& 98.20& 90.66\\ \hline
M4                & 90.12& 88.41& 98.26& 93.07\\ \hline
\textbf{M5}& \textbf{98.88}& \textbf{99.53}& \textbf{98.99}&  \textbf{99.26}\\ \hline
\end{tabular}
\caption{Performance of unimodal and multimodal models under adversarial examples (\%).}
\label{tab:adv-comparison}
\end{table}

\subsubsection{RQ4}

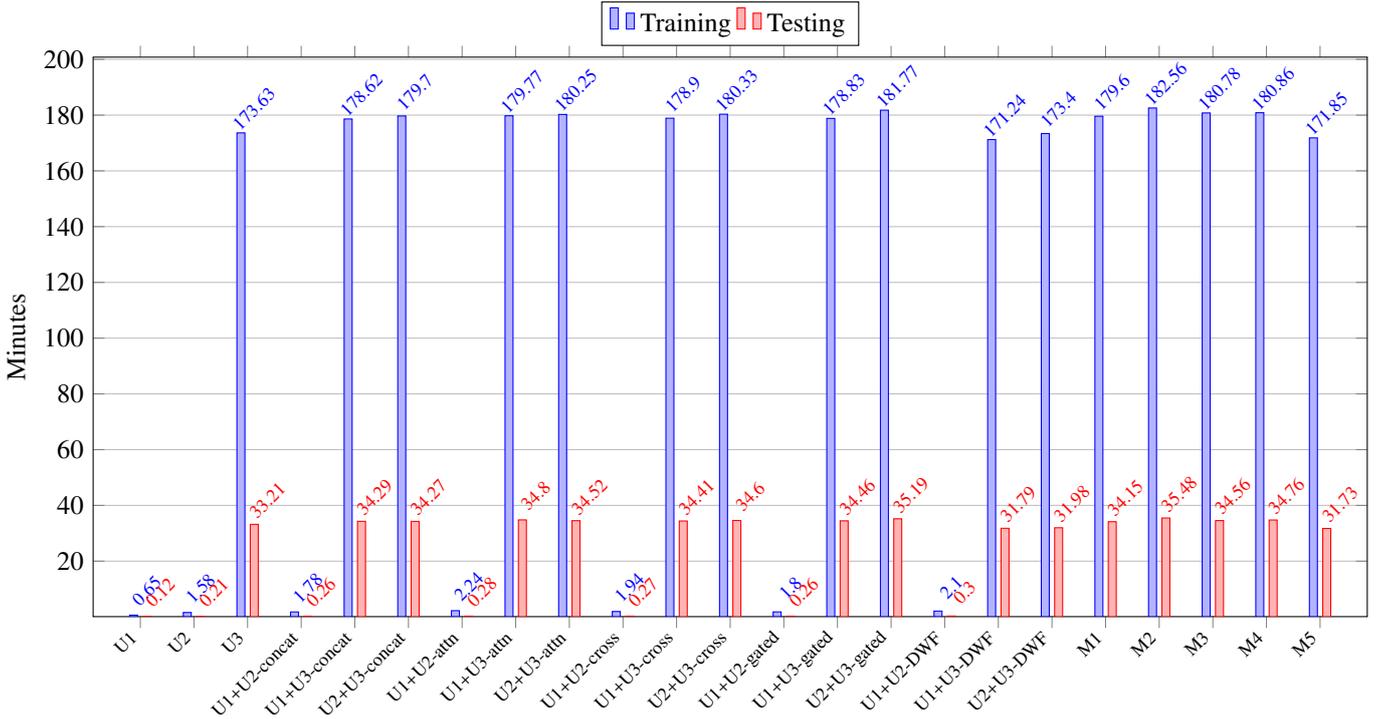
\begin{figure*}[!ht]
\centering
\begin{tikzpicture}
\begin{axis}[
    ybar,
    bar width=3pt,
    width=\textwidth,
    height=9cm,
    enlarge x limits=0.04,
    ymin=0.1,
    ylabel={Minutes},
    nodes near coords,
    every node near coord/.append style={
      font=\scriptsize,
      rotate=45,
      xshift=10pt,yshift=-3pt,
    },
    symbolic x coords={
      {U1},
      {U2},
      {U3},
      {U1+U2-concat},
      {U1+U3-concat},
      {U2+U3-concat},
      {U1+U2-attn},
      {U1+U3-attn},
      {U2+U3-attn},
      {U1+U2-cross},
      {U1+U3-cross},
      {U2+U3-cross},
      {U1+U2-gated},
      {U1+U3-gated},
      {U2+U3-gated},
      {U1+U2-DWF},
      {U1+U3-DWF},
      {U2+U3-DWF},
      {M1},
      {M2},
      {M3},
      {M4},
      {M5}
    },
    xtick=data,
    x tick label style={rotate=45, anchor=east, font=\scriptsize},
    ymajorgrids=true,
    legend style={at={(0.5,1.02)}, anchor=south, legend columns=-1},
    legend cell align=left
]
  \addplot coordinates {
    ({U1},0.65)
    ({U2},1.58)
    ({U3},173.63)
    ({U1+U2-concat},1.78)
    ({U1+U3-concat},178.62)
    ({U2+U3-concat},179.70)
    ({U1+U2-attn},2.24)
    ({U1+U3-attn},179.77)
    ({U2+U3-attn},180.25)
    ({U1+U2-cross},1.94)
    ({U1+U3-cross},178.90)
    ({U2+U3-cross},180.33)
    ({U1+U2-gated},1.80)
    ({U1+U3-gated},178.83)
    ({U2+U3-gated},181.77)
    ({U1+U2-DWF},2.10)
    ({U1+U3-DWF},171.24)
    ({U2+U3-DWF},173.40)
    ({M1},179.60)
    ({M2},182.56)
    ({M3},180.78)
    ({M4},180.86)
    ({M5},171.85)
  };

  \addplot coordinates {
    ({U1},0.12)
    ({U2},0.21)
    ({U3},33.21)
    ({U1+U2-concat},0.26)
    ({U1+U3-concat},34.29)
    ({U2+U3-concat},34.27)
    ({U1+U2-attn},0.28)
    ({U1+U3-attn},34.80)
    ({U2+U3-attn},34.52)
    ({U1+U2-cross},0.27)
    ({U1+U3-cross},34.41)
    ({U2+U3-cross},34.60)
    ({U1+U2-gated},0.26)
    ({U1+U3-gated},34.46)
    ({U2+U3-gated},35.19)
    ({U1+U2-DWF},0.30)
    ({U1+U3-DWF},31.79)
    ({U2+U3-DWF},31.98)
    ({M1},34.15)
    ({M2},35.48)
    ({M3},34.56)
    ({M4},34.76)
    ({M5},31.73)
  };

  \legend{Training, Testing}
\end{axis}
\end{tikzpicture}
\caption{Computational cost per model.}
\label{fig:time-consumption}
\end{figure*}

Training and testing times of all modalities are reported in \textbf{Figure \ref{fig:time-consumption}}, measured over 30 epochs of training. Across all settings, runtime is dominated by the U3 encoder. A single-modality U3 model requires 173.63 minutes for training and 33.21 minutes for testing, whereas the U1 and U2 models are extremely lightweight, training within 0.65–1.58 minutes and testing within 0.12–0.21 minutes. Bimodal fusions that exclude U3 remain efficient, requiring only 1.78–2.24 minutes to train and 0.26–0.30 minutes to test. However, any configuration that includes U3 inherits its high computational cost, with training times between 171.24 and 182.56 minutes and testing times between 31.73 and 35.48 minutes. Concretely, the M4 increases inference time by about 134 times over U1+U2-gated (34.76 vs 0.26 minutes) and about 290 times over U1 (34.76 vs 0.12 minutes), and increases training time by about 102 times over U1+U2-gated (180.86 vs 1.78 minutes). In contrast, U1+U2-gated itself is only approximately 2.7 times slower to train and approximately 2.2 times slower to test than U1, while being far cheaper than any U3-based model.

Despite its higher computational overhead, U3 contributes significantly to multimodal robustness against obfuscation and adversarial attacks. Incorporating U3 even in bimodal settings substantially improves robustness compared to U1-based pairs without it, and in the trimodal case, U3 enables M5 with DWF to reach near-clean performance. Overall, these results highlight a clear trade-off between detection performance and computational cost. Lightweight models such as U1, U2, or their fusion achieve efficient training and testing but provide lower robustness, whereas U3-based models, while computationally expensive, deliver substantially higher detection reliability and resistance to sophisticated evasion strategies.

\subsubsection{Comparison experiment}
We re-implement two representative multimodal approaches (X. Li et al. \cite{li2025detecting} and Chimera \cite{de2023chimera}), as discussed in \textbf{Section \ref{sec_relatedworks}}, and evaluate their performance on the CICMalDroid 2020 dataset. The evaluation is also conducted on the test set, covering original samples as well as their obfuscated and adversarial samples to ensure a comprehensive and fair comparison under diverse conditions. Among the two baseline approaches, only the method by \cite{li2025detecting} provides open-source code, whereas Chimera does not release its implementation. For Chimera, we relied on the descriptions in the original paper, which were based on the Omnidroid dataset. Since our re-implementation uses CICMalDroid 2020, discrepancies arise in certain features such as system calls and permissions \& intents, potentially leading to differences in performance. 

The comparative results between these two multimodal approaches and our proposed method are summarized in \textbf{Table \ref{tab:comparison}}. On the original test set, all three methods demonstrate high efficacy, with our approach achieving an F1-score of 98.67, X. Li et al. reaching 98.69, and Chimera attaining 95.01, establishing a strong baseline for detection on clean data.

However, these strong results weaken once the models are subjected to evasion techniques. Under adversarial conditions, DMLDroid achieves the highest F1-score of 99.26 and a recall of 99.53, demonstrating that its dynamic weighted fusion effectively balances contributions from all modalities, mitigating the impact of targeted perturbations. X. Li et al. also perform well with an F1-score of 98.25 because the attack targets permissions and intents, which their model does not use. In contrast, Chimera drops to 90.79 in accuracy and 93.67 in F1-score due to its reliance on permissions and intents and its simple concatenation fusion, which cannot compensate for compromised modalities.

This pattern of superior resilience is further confirmed across the obfuscation scenarios, where the architectural weaknesses of the baseline models become even more apparent. Under code obfuscation (Co), both baseline models suffer severe degradation. Specifically, the F1-score of X. Li et al. drops to 83.91, a decline of more than 14\% compared to its original F1-score of 98.69. This decline arises because its fusion strategy depends heavily on fine-grained bytecode features, which are highly susceptible to syntactic transformations. Techniques such as \textit{Goto} and CallIndirection scramble opcode sequences and obscure API calls, effectively collapsing the discriminative power of its representation. Similarly, Chimera experiences an even sharper performance breakdown, with its F1-score plunging to 74.94 from 95.01. This is because its grayscale image representation lacks any structural semantics, causing syntactic changes in the bytecode to create major variations, making the visual patterns learned by the model unrecognizable.

The weakness of these approaches becomes even more apparent under mixed obfuscation (Rn+Co+Enc). When renaming, code obfuscation, and encryption are jointly applied, X. Li et al. achieve an F1-score of only 81.77, while Chimera degrades further to 74.76. These results confirm that both methods overfit to modality-specific syntactic signatures, and their fusion mechanisms lack the flexibility to adapt when one modality is compromised.

In contrast, DMLDroid demonstrates remarkable robustness, maintaining an F1-score of 99.29 under Co and a still exceptional 98.95 under the combined Rn+Co+Enc scenario. This resilience stems from its dynamic weighted fusion mechanism, which adaptively adjusts the confidence scores assigned to each modality, thereby reducing over-dependence on any modality that becomes compromised.

\begin{table*}[!ht]
\centering
\small
\caption{Comparative results of different fusion-based models across evaluation scenarios.}
\label{tab:comparison}
\begin{tabular}{|cc|cccc|cccc|cccc|}
\hline
\multicolumn{2}{|c|}{}                                                      & \multicolumn{4}{c|}{\textbf{X. Li et al. \cite{li2025detecting}}}                                                                              & \multicolumn{4}{c|}{\textbf{Chimera \cite{de2023chimera}}}                                                                                   & \multicolumn{4}{c|}{\textbf{DMLDroid (Ours)}}                                                                                                                                                           \\ \cline{3-14} 
\multicolumn{2}{|c|}{\multirow{-2}{*}{\textbf{Test scenario}}}              & \multicolumn{1}{c|}{\textbf{Acc}} & \multicolumn{1}{c|}{\textbf{Rec}} & \multicolumn{1}{c|}{\textbf{Pre}} & \textbf{F1} & \multicolumn{1}{c|}{\textbf{Acc}} & \multicolumn{1}{c|}{\textbf{Rec}} & \multicolumn{1}{c|}{\textbf{Pre}} & \textbf{F1} & \multicolumn{1}{c|}{\textbf{Acc}}                  & \multicolumn{1}{c|}{\textbf{Rec}}                  & \multicolumn{1}{c|}{\textbf{Pre}}                  & \textbf{F1}                   \\ \hline
\multicolumn{2}{|c|}{Original test}                                         & \multicolumn{1}{c|}{98.10}        & \multicolumn{1}{c|}{98.56}        & \multicolumn{1}{c|}{98.81}        & 98.69       & \multicolumn{1}{c|}{92.57}        & \multicolumn{1}{c|}{93.24}        & \multicolumn{1}{c|}{96.86}        & 95.01       & \multicolumn{1}{c|}{97.98} & \multicolumn{1}{c|}{98.32} & \multicolumn{1}{c|}{99.02} & 98.67 \\ \hline
\multicolumn{2}{|c|}{Adversarial}                                           & \multicolumn{1}{c|}{\textbf{97.51}}        & \multicolumn{1}{c|}{\textbf{97.75}}        & \multicolumn{1}{c|}{\textbf{98.75}}        & \textbf{98.25}       & \multicolumn{1}{c|}{\textbf{90.79}}          & \multicolumn{1}{c|}{\textbf{90.89}}          & \multicolumn{1}{c|}{\textbf{96.63}}          & \textbf{93.67}         & \multicolumn{1}{c|}{\textbf{98.88}}                         & \multicolumn{1}{c|}{\textbf{99.53}}                         & \multicolumn{1}{c|}{\textbf{98.99}}                         & \textbf{99.26}                         \\ \hline
\multicolumn{1}{|c|}{}                              & Rn         & \multicolumn{1}{c|}{98.28}        & \multicolumn{1}{c|}{98.87}        & \multicolumn{1}{c|}{98.66}        & 98.77       & \multicolumn{1}{c|}{82.81}        & \multicolumn{1}{c|}{95.91}        & \multicolumn{1}{c|}{80.05}        & 87.27       & \multicolumn{1}{c|}{98.93}                         & \multicolumn{1}{c|}{99.64}                         & \multicolumn{1}{c|}{98.91}                         & 99.27                         \\ \cline{2-14} 
\multicolumn{1}{|c|}{}                              & Co & \multicolumn{1}{c|}{\textbf{79.63}}        & \multicolumn{1}{c|}{\textbf{73.12}}        & \multicolumn{1}{c|}{\textbf{98.43}}        & \textbf{83.91}       & \multicolumn{1}{c|}{\textbf{69.18}}        & \multicolumn{1}{c|}{\textbf{61.88}}        & \multicolumn{1}{c|}{\textbf{95.00}}        & \textbf{74.94}       & \multicolumn{1}{c|}{\textbf{98.94}}                         & \multicolumn{1}{c|}{\textbf{99.63}}                         & \multicolumn{1}{c|}{\textbf{98.96}}                         & \textbf{99.29}                         \\ \cline{2-14} 
\multicolumn{1}{|c|}{}                              & Enc      & \multicolumn{1}{c|}{98.54}        & \multicolumn{1}{c|}{99.21}        & \multicolumn{1}{c|}{98.73}        & 98.97       & \multicolumn{1}{c|}{88.92}          & \multicolumn{1}{c|}{88.36}          & \multicolumn{1}{c|}{96.27}          & 92.14         & \multicolumn{1}{c|}{99.01}                         & \multicolumn{1}{c|}{99.72}                         & \multicolumn{1}{c|}{98.96}                         & 99.34                         \\ \cline{2-14} 
\multicolumn{1}{|c|}{\multirow{-4}{*}{Obfuscation}} & Rn+Co+Enc             & \multicolumn{1}{c|}{\textbf{77.65}}        & \multicolumn{1}{c|}{\textbf{70.01}}        & \multicolumn{1}{c|}{\textbf{98.28}}        & \textbf{81.77}       & \multicolumn{1}{c|}{\textbf{69.38}}          & \multicolumn{1}{c|}{\textbf{61.74}}          & \multicolumn{1}{c|}{\textbf{94.72}}          & \textbf{74.76}         & \multicolumn{1}{c|}{\textbf{98.50}}                         & \multicolumn{1}{c|}{\textbf{99.12}}                         & \multicolumn{1}{c|}{\textbf{98.79}}                         & \textbf{98.95}                         \\ \hline
\end{tabular}
\end{table*}

\section{Threats to validity} \label{sec_threat_to_validity}
\subsection{External validity}
The external validity of our study may be affected by dataset dependency and evaluation scope. Although the proposed method shows strong results on CICMalDroid 2020, generalization to other datasets (e.g., Drebin, AMD, AndroZoo) or to novel malware families remains uncertain. Our evaluation also relies exclusively on static features (permissions, intents, DEX images, API sequences), which may not capture dynamic or hybrid behaviors observed in real-world execution environments. Furthermore, robustness analysis was restricted to obfuscation generated by Obfuscapk, while commercial or layered obfuscation frameworks (e.g., ProGuard, DexGuard, Allatori, DashO) might produce more sophisticated variants. These factors collectively limit the generalizability of our findings across broader and evolving malware ecosystems.

\subsection{Internal validity}

Internal validity may be influenced by model interpretability and assumptions in the fusion mechanism. While Dynamic Weighted Fusion (DWF) consistently improves robustness, the decision process remains opaque, as no explainability methods (e.g., SHAP, LIME, or modality-level attribution) were applied to identify which features drive resilience. This raises the risk that performance gains stem from dataset-specific artifacts rather than meaningful malware semantics. Moreover, DWF implicitly assumes that at least two modalities remain reliable under attack. Our evaluation did not consider scenarios where multiple modalities are simultaneously compromised, which could undermine the weighting mechanism. These limitations highlight the need for interpretability-driven analysis and stress-testing under more severe adversarial conditions.

\subsection{Construct validity}

A potential limitation lies in how “robustness” is defined and measured. This study relies solely on Accuracy, Precision, Recall, and F1-score to evaluate resilience; other important aspects, such as detection latency, computational overhead, and interpretability, have not been analyzed. The choice of three feature types (permissions \& intents, DEX images, API sequences) reflects common research directions but may not fully capture malware behaviors, for example, dynamic system calls or network traffic. In addition, the tools used to generate obfuscation (Obfuscapk) and adversarial examples (GAN-based perturbations on permissions/intents) may introduce bias, as they do not cover all real-world attack techniques. Therefore, the construct validity of “adversarial resilience” is contingent on the scope of definitions and implementations in this study.

\subsection{Conclusion validity}
The conclusions drawn about the effectiveness of multimodal fusion and Dynamic Weighted Fusion (DWF) are primarily based on a single dataset (CICMalDroid 2020) with a fixed train/test split. The absence of evaluations on multiple independent datasets (e.g., Drebin, AMD, AndroZoo) or cross-validation may limit the statistical reliability of the findings. Moreover, small performance differences between models (e.g., 0.5–1\% in F1-score) have not been verified with statistical significance tests (e.g., t-test, confidence intervals), making it difficult to claim their practical importance. Additionally, the results on computational cost (\textbf{RQ4}) were obtained in a fixed GPU environment; in real-world deployments, variations in hardware or Android operating system versions could lead to different outcomes. These factors should be considered when generalizing and interpreting the conclusions.

\section{Conclusion} \label{sec_conclusion}
This paper presents DMLDroid, an Android malware detection framework that employs multimodal intermediate fusion to achieve strong resilience against both obfuscation and adversarial attacks. The framework integrates three complementary feature types: tabular features derived from permissions and intents, visual features extracted from DEX file representations, and sequential features obtained from API call graphs. In the pre-training phase, modality-specific representations are first learned to build a high-quality initial model that is inherently robust to adversarial perturbations and code obfuscation. This process supports fine-tuning the language backbone (DistilBERT) and enhances convergence and generalization in CNN and MLP training. We deliberately include permissions and intents as a tabular modality, despite their vulnerability to adversarial manipulations, to investigate how different fusion mechanisms adapt and reallocate representational capacity when one modality is degraded or compromised. Accordingly, we compare five intermediate fusion strategies, including concatenation, self-attention, cross-attention, gated fusion, and dynamic weighted fusion, and find that dynamic weighted fusion consistently delivers the most robust performance across all scenarios. In terms of efficiency, the testing phase takes 31.73 minutes for over 5000 samples, corresponding to only 0.38 seconds per sample, demonstrating both resilience and practical inference speed. 

While the study shows strong robustness and efficiency, it is not without limitations, as discussed in \textbf{Section \ref{sec_threat_to_validity}}. Future work will extend evaluations to cases where multiple modalities are compromised, validate on more diverse datasets, and explore lighter sequence models to reduce computational cost.


\section*{Acknowledgment}

This research is funded by Vietnam National University Ho Chi Minh City (VNU-HCM) under grant number NCM2025-26-01.




\bibliographystyle{elsarticle-num} 
\bibliography{references}





\end{document}